\documentclass[10pt,journal,compsoc]{IEEEtran}
\usepackage{amsmath,amssymb}
\usepackage{algpseudocode}
\usepackage{enumitem}
\usepackage{graphicx}
\usepackage{makecell} %
\usepackage{multirow}
\usepackage{pifont} %

\usepackage{stfloats} %

\usepackage{stmaryrd} %
\usepackage{threeparttable} %
\usepackage{xspace}
\usepackage{textcomp} %

\usepackage[colorlinks,
 linkcolor=black,
 anchorcolor=black,
 citecolor=black
]{hyperref}

\def\ie{\textit{i.e.}\xspace}

\def\eg{\textit{e.g.}\xspace}
\def\etal{\textit{et~al.}\xspace}
\def\ds{\mathcal{U}\xspace}
\def\sone{\mathcal{S}_1\xspace}
\def\stwo{\mathcal{S}_2\xspace}

\newcommand{\pk}{\mathit{pk}}
\newcommand{\sk}{\mathit{sk}}

\pdfnoligatures\font
\emergencystretch=\maxdimen
\begin{document}

\title{Optimizing Privacy-Preserving Outsourced Convolutional Neural Network Predictions}

\author{Minghui~Li,
Sherman S. M. Chow,
Shengshan~Hu,
Yuejing~Yan,
Chao~Shen,
and Qian~Wang%
\IEEEcompsocitemizethanks{\IEEEcompsocthanksitem M. Li, and Q. Wang are with the School of Cyber Science and Engineering, School of Computer Science, Wuhan University, Wuhan 430072, Hubei, China, and State Key Laboratory of Cryptology, P.O. Box 5159, Beijing, 100878, China. \protect
E-mail: \{minghuili, qianwang\}@whu.edu.cn
\IEEEcompsocthanksitem S. Chow is with the Department of Information Engineering, The Chinese University of Hong Kong, Hong Kong. \protect E-mail: sherman@ie.cuhk.edu.hk
\IEEEcompsocthanksitem S. Hu is with the Services Computing Technology and System Laboratory, Cluster and Grid Computing Laboratory, National Engineering Research Center for Big Data Technology and System,  School of Cyber Science and Engineering, Huazhong University of Science and Technology, Wuhan 430074, Hubei, China. \protect E-mail: hushengshan@hust.edu.cn
\IEEEcompsocthanksitem Y. Yan is with State Key Laboratory of Information Engineering in Surveying, Mapping, and Remote Sensing, Wuhan University, Wuhan 430072, Hubei, China. \protect E-mail: yjyan@whu.edu.cn
\IEEEcompsocthanksitem C. Shen is with School of Cyber Security, School of Electronic and Information Engineering, Xi'an Jiaotong University, Xian 710049, Shaanxi, China. \protect E-mail: cshen@sei.xjtu.edu.cn
 }}

\IEEEtitleabstractindextext{%
\begin{abstract}
Convolutional neural network is a machine-learning model widely applied in various prediction tasks, such as computer vision and medical image analysis.
Their great predictive power requires extensive computation, which encourages model owners to host the prediction service in a cloud platform.
Recent researches focus on the privacy of the query and results, but they do not provide model privacy against the model-hosting server and may leak partial information about the results.
Some of them further require frequent interactions with the querier or heavy computation overheads, which discourages querier from using the prediction service.
This paper proposes a new scheme for privacy-preserving neural network prediction in the outsourced setting, \ie, the server cannot learn the query, (intermediate) results, and the model.
Similar to SecureML (S\&P'17), a representative work that provides model privacy, we leverage two non-colluding servers with secret sharing and triplet generation to minimize the usage of heavyweight cryptography.
Further, we adopt asynchronous computation to improve the throughput, and design garbled circuits for the non-polynomial activation function to keep the same accuracy as the underlying network (instead of approximating it).
Our experiments on MNIST dataset show that our scheme achieves an average of $122\times$, $14.63\times$, and $36.69\times$ reduction in latency compared to SecureML, MiniONN (CCS'17), and EzPC (EuroS\&P'19), respectively.
For the communication costs, our scheme outperforms SecureML by $1.09\times$, MiniONN by $36.69\times$, and EzPC by $31.32\times$ on average.
On the CIFAR dataset, our scheme achieves a lower latency by a factor of $7.14\times$ and $3.48\times$ compared to MiniONN and EzPC, respectively.
Our scheme also provides $13.88\times$ and $77.46\times$ lower communication costs than MiniONN and EzPC on the CIFAR dataset.
\end{abstract}

}

\maketitle

\IEEEdisplaynontitleabstractindextext

\IEEEpeerreviewmaketitle

\section{Introduction}

\IEEEPARstart{M}{achine} learning (ML)~\cite{Goodfellow-et-al-2016} performs well in many applications and has been widely used (\eg,~\cite{tifs/ZhaoWZZC19,tifs/ZouWWZL20,tvt/ZouJDYCW20}).
Neural networks, which identify relationships underlying a set of data by mimicking how the human brain operates, have recently gained extensive attention.
Convolutional neural networks (CNN), based on biologically-inspired variants of multi-layer perceptrons, are proven to be useful in medical image analysis and recognition of images and videos.

With the popularity of machine-learning-as-a-service (MLaaS), the model owners tend to host the model in the cloud-based MLaaS for providing prediction services.
Nevertheless, it is tempting for an adversary to steal the model~\cite{uss/TramerZJRR16}, 
pirate it, or use it to provide a commercial prediction service for profits~\cite{uss/AdiBCPK18} 
since it is a valuable asset.
Moreover, with the knowledge of the model, the risk of compromising the model privacy is higher since white-box attacks can infer more information than black-box attacks,
\eg, membership inference attack~\cite{sp/ShokriSSS17,popets/HayesMDC19,ndss/Salem0HBF019} for determining if a specific sample was in the training dataset.
Knowing the model also makes adversarial example attacks~\cite{ccs/PapernotMGJCS17,uss/JiaG18,nips/ElsayedSCPKGS18, tdsc/WangSZZSW19} more effective.
A tiny perturbation of the input can deceive the model and threaten its accuracy.

Outsourcing the model and the prediction service to any untrusted cloud platform (say, for relieving from the cost of maintaining an online server)
thus comes with great privacy and security implications.
This paper aims to propose an efficient prediction service that ensures \emph{model privacy}, \ie,
keeping the model private from the querier (as most existing works) and any hosting server.

\subsection{Related Work}
Techniques in preserving privacy in neural network prediction can be broadly categorized into differential privacy, trusted processors, and cryptography. 
Differential privacy~\cite{AbadiCGMMT016,SmithTU17} adds noise without sacrificing too much data utility, but it cannot ensure data privacy as much as the other two classes of techniques.
Trusted processor (\eg, SGX) approaches~\cite{OhrimenkoSFMNVC16,tdsc/HuZWQW19,iclr/TramerB19,scc/Chow19,ng2020goten} work on the data within the trusted perimeter,
but they are subjected to the memory constraint (currently $128$MB) which can be easily exceeded by a specific layer of a deep neural network.
Cryptographic approaches do not have these problems but with higher overheads. 
Ensuring privacy with a tailored cryptographic design is a recurrent research problem.

Many existing schemes protect the privacy of the \emph{query} from the server.
Yet, they consider the model owner is the server performing the prediction tasks, \ie, the prediction service only works with the \emph{plaintext knowledge of the model}.
Some works~\cite{Gilad-BachrachD16,ChabanneWMMP17} use additive homomorphic encryption to perform operations over the encrypted query and the clear model.
Others~\cite{LiuJLA17,uss/JuvekarVC18,dac/RouhaniRK18,uss/RiaziSCLLK19,eurosp/ChandranGRST19} design secure two-party computation (S2C) protocols for various kinds of machine-learning computations.
MiniONN~\cite{LiuJLA17} and Gazelle~\cite{uss/JuvekarVC18} adopted secret sharing in which the query is secret-shared (between the user) with the server.
DeepSecure~\cite{dac/RouhaniRK18} preprocesses the data and the neural network before
S2C but leaks some information about the parameters.
XONN~\cite{uss/RiaziSCLLK19} ``replaces'' the matrix multiplication with XNOR, which is virtually free in GC.
Yet, this scheme only applies to binary neural networks, \ie, the parameters are binary.
EzPC~\cite{eurosp/ChandranGRST19} proposes a compiler that translates between arithmetic and boolean circuits.
However, the S2C-based approach often expects the queriers to remain online and interact with the server continuously, thus bringing some burden to the queriers and incurring higher network round-trip time.

Fully-homomorphic encryption (FHE)~\cite{gentry09} allows processing (any polynomial functions over) encrypted data.
It is thus a handy tool for not only processing the query in an encrypted format
but also processing over an encrypted model.
Bost~\etal~\cite{ndss/BostPTG15} considered various machine-learning classifications (\eg, hyperplane decision-based classifiers, na\"ive Bayes classifiers, decision trees) over FHE-encrypted data.
In the case of decision trees,
Tai~\etal~\cite{esorics17/TaiMZC} managed to use only additive HE
instead of FHE to support S2C evaluation,
but it is subject to the limitation that the server needs to know the model in clear.
Using multi-key FHE, Aloufi~\etal~\cite{tdsc/AloufiHWC19} considered secure outsourcing of decision-trees evaluation.
The evaluation results can only be decrypted with the help of multiple secret keys from multiple parties.
Chow~\cite{fcs/Chow18} provided a brief overview of the state-of-the-art in privacy-preserving decision tree evaluation.
All these works did not consider neural networks. 
Indeed, processing an FHE-encrypted neural network directly without optimization is time-consuming.
FHE also fails to cover common non-polynomial operations in neural networks.
CryptoDL~\cite{HesamifardTGW18} and E2DM~\cite{ccs/JiangKLS18} approximate them by polynomials, which degrades the prediction accuracy.
A major drawback is that these approaches require the model owner to encrypt the model w.r.t the public key of each querier.
It not only does not scale when there are multiple queriers,
but also increases the risk of model leakage since any querier has the power to decrypt the model.

\subsection{Two-Server Computation Model}
To reduce the use of cryptographic techniques, many works~\cite{ndss/ChowLS09,iacr/KamaraMR11,cns/WangLC014,BaryalaiJL16,tkde/WangDCCZCH18} exploited the non-colluding assumption 
for the possibility 
of using lightweight cryptographic primitives, 
such as replacing the use of FHE with additive HE~\cite{cns/WangLC014,BaryalaiJL16}.
SecureML~\cite{MohasselZ17} uses \emph{additive secret sharing}~\cite{cacm/Shamir79}
to share the model 
among two servers.
The querier also secret-shares the query across the servers.
To carry out prediction, the two servers interact with each other and
eventually derive their corresponding share of the prediction result.
The querier can then obtain the final result by merging both shares locally.
The benefit of using secret sharing is that both servers can operate over the secret share, without knowing the underlying secret, almost as efficient as the operating over the secret itself.
When the servers do not collude, none of the respective shares reveals anything about the secret.

To speed up the online (inner-product) computation,
SecureML generates Beaver's multiplication \emph{triplet}~\cite{Beaver91a} for additive sharing in an offline preprocessing phase.
For the non-polynomial operation (\ie, comparison), they use Yao's garbled circuits (GC)~\cite{Yao86} for boolean computations.
Yet, SecureML only focuses on simple neural networks (and linear/logistic regression)
without considering the more complex convolutional neural networks.

Looking ahead, our scheme follows this design 
and incorporates efficiency improvement over it.
Fig.~\ref{fig:system_model} overviews such a design, to be explained in Section~\ref{subsec:system_model}.

Table~\ref{tab:Leakeage} coarsely compares existing schemes.
It is fair to say that there is no outsourcing solution with model privacy
and satisfactory performance in accuracy and efficiency.

\begin{table}[!t]
\setlength{\belowcaptionskip}{3pt}
\centering
\newcolumntype{M}[1]{>{\centering\arraybackslash}m{#1}}
\renewcommand{\arraystretch}{1.6}
\addtolength{\tabcolsep}{-0pt}
\caption{Comparison of related work}

\begin{tabular}{ M{7em}||M{2.3em}|M{2.1em}|M{2.2em}||M{3.8em}|M{3.8em} }
 \Xhline{1.1pt}
\multirow{2}{*}{} &\multicolumn{3}{c||}{Privacy} &\multirow{3}{*}{Accuracy} &\multirow{3}{*}{Efficiency} \\
\cline{2-4}
 &model para. &inter. data & query & \\
 \Xhline{1.1pt}

CryptoNets~\cite{Gilad-BachrachD16}&\ding{55} & \ding{51}& \ding{51}& Low &Medium\\
 \Xhline{0.7pt}

CryptoDL~\cite{HesamifardTGW18}&\ding{51} & \ding{51}& \ding{51}&Low &Medium\\ %
 \Xhline{0.7pt}

E2DM~\cite{ccs/JiangKLS18}&\ding{51} & \ding{51}& \ding{51}&Low &Medium\\
 \Xhline{0.7pt}

 XONN~\cite{uss/RiaziSCLLK19}&\ding{55} & \ding{51}& \ding{51}&Medium &High\\
 \Xhline{0.7pt}

 DeepSecure~\cite{dac/RouhaniRK18}&\ding{55} & \ding{51}& \ding{51}&High &Medium\\
 \Xhline{0.7pt}

SecureML~\cite{MohasselZ17}&\ding{51} & \ding{51}& \ding{51}& High&Medium\\
 \Xhline{0.7pt}

MiniONN~\cite{LiuJLA17}&\ding{55} & \ding{51}& \ding{51}&High &High\\
 \Xhline{0.7pt}

Gazelle~\cite{uss/JuvekarVC18}&\ding{55} & \ding{51}& \ding{51}& High&High\\
 \Xhline{0.7pt}

 EzPC~\cite{eurosp/ChandranGRST19}&\ding{55} & \ding{51}& \ding{51}&High &High\\
 \Xhline{0.7pt}

Our Scheme&\ding{51} & \ding{51}& \ding{51}& High&High\\
 \Xhline{1.1pt}
\end{tabular}
\label{tab:Leakeage}
\end{table}

\subsection{Our Contributions}
proposes a new scheme that can simultaneously protect the query, the model, any intermediate results, and the final prediction results against the servers.
Our contributions lie in maintaining high accuracy and efficiency, which are summarized below.

\begin{itemize}[leftmargin=10pt]
	\item We design protocols for different stages of the convolutional neural network prediction.
	\item %
	We accelerate the triplets generation~\cite{MohasselZ17,Beaver91a} using 
	single-instruction-multiple-data (SIMD)~\cite{dcc/SmartV14}.
	We also adopt asynchronous computation to speed up both offline and online computations.
	\item For non-polynomial activation functions, we design a series of garbled circuits for S2C between the two servers.
	In contrast to existing approaches, which only approximate the activation functions by polynomials,
	our approach preserves the accuracy.
 	\item
	We also replace the non-polynomial max-pooling function with the average-pooling function, 
	which is a linear function for further efficiency improvement.
	In our experiment, we train with the MNIST dataset and CIFAR dataset. Our results demonstrate that the final accuracy of average-pooling is comparable to that of max-pooling.
	\item Our experiments on MNIST dataset show that our scheme achieves $122\times$, $14.63\times$, and $8.19\times$ lower latency than SecureML~\cite{MohasselZ17}, MiniONN~\cite{LiuJLA17}, and EzPC~\cite{eurosp/ChandranGRST19}, respectively.
	For the communication costs, our scheme outperforms 
	SecureML by $1.09\times$, MiniONN by $36.69\times$, and EzPC by $31.32\times$.
	Our scheme also incurs $7.14\times$ and $3.48\times$ lower computation costs, and $13.88\times$ and $77.46\times$ lower communication costs, than MiniONN and EzPC on the CIFAR dataset.
	\item We provide a security analysis in the simulation paradigm to show the privacy of our scheme.
\end{itemize}

\section{Preliminary}
\subsection{Convolutional Neural Network (CNN) and Notations}

\begin{table}[!t]
\setlength{\belowcaptionskip}{3pt}
\centering
\newcolumntype{M}[1]{>{\centering\arraybackslash}m{#1}}
\renewcommand{\arraystretch}{1.6}
\addtolength{\tabcolsep}{-0pt}
\caption{Notations}
\begin{tabular}{ M{4em}||M{20em}}
\Xhline{1.1pt}
$\mathsf{C}$ & ($m \times m$) convolutional kernels of CNN\\
\hline
$f(x)$ & non-polynomial activation function\\
\hline
$\mathcal{G}$ & original image (of pixel size $n\times n$)\\
\hline
$\mathcal{H}$ & convoluted image\\
\hline
$\mathcal{J}$ & activated image\\
\hline
$q\times q$ & size of the pooling window\\
\hline
$\mathcal{K}$ & down-sampled image\\
\hline
$\mathsf{W}$ & weight matrix in fully-connected layer\\
\hline
$\mathcal{F}$ & prediction results\\
\hline
$\cdot$ & inner product operation\\
\Xhline{1.1pt}
\end{tabular}
\label{tab:notation}
\end{table}

We use image processing as a running example since CNN is good at such kind of tasks.
A typical CNN consists of four classes of layers.
The core one is the \emph{convolutional layer}.
It computes the convolution, essentially an \emph{inner product}, between the raw pixel values in each local region of the image and a set of the convolutional kernel.
Table~\ref{tab:notation} lists our major notation.
For the inner product (denoted by~$\cdot$) of two matrices, we first transform the matrices into vectors.

Let $\mathcal{G}$ be the query image as an $n \times n$ matrix of $8$-bit pixel.
For simplicity, we consider only one convolution kernel $\mathsf{C}$ each of size $m \times m$ (\ie, the padding mode is ``same'').
The convolution transforms $\mathcal{G}$ into a matrix~$\mathcal{H}$.

After the convolutional layer, it is usually the \emph{activation layer}, which applies an element-wise \emph{non-polynomial} activation function for increasing the nonlinear properties of the model.
We use the most fashionable ReLU function, which only keeps the positive part of any real-valued input (\ie, $f(x) = \max(0,x)$).
We let $\mathcal{J}$ be the image activated by $f(x)$, which remains of size $n \times n$.

The \emph{pooling layer} performs a down-sampling along the spatial dimensions while retaining critical information.
The usual ones are \emph{max-pooling} and \emph{average-pooling}, which outputs the maximum or the average value of the pool.
The size of the pool is $q\times q$.
The resulting smaller image $\mathcal{K}$ is of size $(n/q)^2$.

The final layer is the \emph{fully-connected layer}.
The image matrix
in the previous layer is first transformed into a column vector of dimension being the neuron numbers of the first layer in the fully-connected layer, \ie, each element in the vector is the input of the corresponding neuron.
Every neuron in the previous layer is connected to every neuron in the next layer.
The weight matrix of the fully-connected layer is denoted by $\mathsf{W}$.
By the scalar product of the output of the previous layer and the corresponding weight matrix, we obtain a column vector, each of its elements corresponds to the input of the next layer.
The output of the last layer is deemed as the final prediction results.

The classes with the desired scores form the final prediction result $\mathcal{F}$, which correspond to the most probable category.
Our goal is to let the querier learn $\mathcal{F}$ without revealing the query $\mathcal{G}$ and the neural network to the servers.

\subsection{Cryptographic Tools}
\subsubsection{Secure Two-Party Computation (S2C)}
S2C allows two parties to jointly compute a function over their private inputs while warranting the correctness.
No party learns anything about the other's input beyond what is implied by the function.
\emph{Garbled circuits} proposed by Yao~\cite{Yao86} can evaluate arbitrary function represented as a boolean circuit.
During the evaluation, no party learns anything about the input data of others beyond what is implied by the final results.

\subsubsection{Secret Sharing}\label{subsec:prelim-secret-sharing}
Many S2C protocols~\cite{ndss/Demmler0Z15} operate on secret-shared inputs.
Secret sharing allows one to distribute a secret by distributing shares to each of many parties.
In $(t, n)$-Shamir secret sharing~\cite{cacm/Shamir79}, there are $n$ shares, which are random elements of a finite field.
Any set of at least $t$ shares allows recovery of the secret value via Lagrange interpolation 
of a degree $(t - 1)$ polynomial embedding the secret.
Any non-qualifying set of shares looks randomly distributed, which provides perfect confidentiality.

In this paper, we use $(2, 2$)-secret sharing
or simply additive secret sharing,
in which the two shares add up to the secret value in the field.
Specifically, consider the secret to be shared is $s$, which is encoded as a $t$-bit string.
One can pick a $t$-bit string $r$ uniformly at random. 
The shares are $r$ and $s - r \bmod{2^t}$.
For sure, $r$ alone is random, and $s - r \bmod{2^t}$ alone is random too 
since, for every candidate $s'$, one must be able to find the corresponding $r'$ 
such that $s' - r' = s - r$ due to the existence of the additive inverse.

We operate on the secret-shared values extensively.
We use a unified notation of superscript $1$ or $2$ to denote the shared information of servers $\sone$ or $\stwo$.
For examples, the shared image held by $\sone$ and $\stwo$ are $\mathcal{G}^1$ and $\mathcal{G}^2$ respectively,
and the $i$-th row and $j$-th column entry held by $\sone$ are $\mathcal{G}_{i,j}^1$.

\subsubsection{Homomorphic Encryption (HE)}
HE (\eg, \cite{gentry09,ima/BosLLN13}) allows computations over encrypted data without access to the secret key.
Decrypting the resultant ciphertexts gives the computation result of the operations as if they had been performed on the plaintext.
For the public/private key pair $(\pk, \sk)$, and an encryption of $x$ denoted by $\llbracket x \rrbracket_{\pk}$, we have homomorphisms $\oplus$ and $\otimes$ where
$\llbracket a \rrbracket_{\pk} \oplus \llbracket b \rrbracket_{\pk} = \llbracket a+b \rrbracket_{\pk}$, $\llbracket a \rrbracket_{\pk} \otimes \llbracket b \rrbracket_{\pk} = \llbracket a\times b \rrbracket_{\pk}$.
Following MiniONN~\cite{LiuJLA17}, this work uses the ring-based FHE scheme called YASHE~\cite{ima/BosLLN13}.
It offers a plaintext space that is large enough for usual computation.
With SIMD~\cite{dcc/SmartV14}, $4096$ independent plaintexts in our application can be packed to one ciphertext and operated in parallel, which reduces the communication and computation.

\begin{figure*}[!ht]
\centering
\includegraphics[width=1.7\columnwidth]{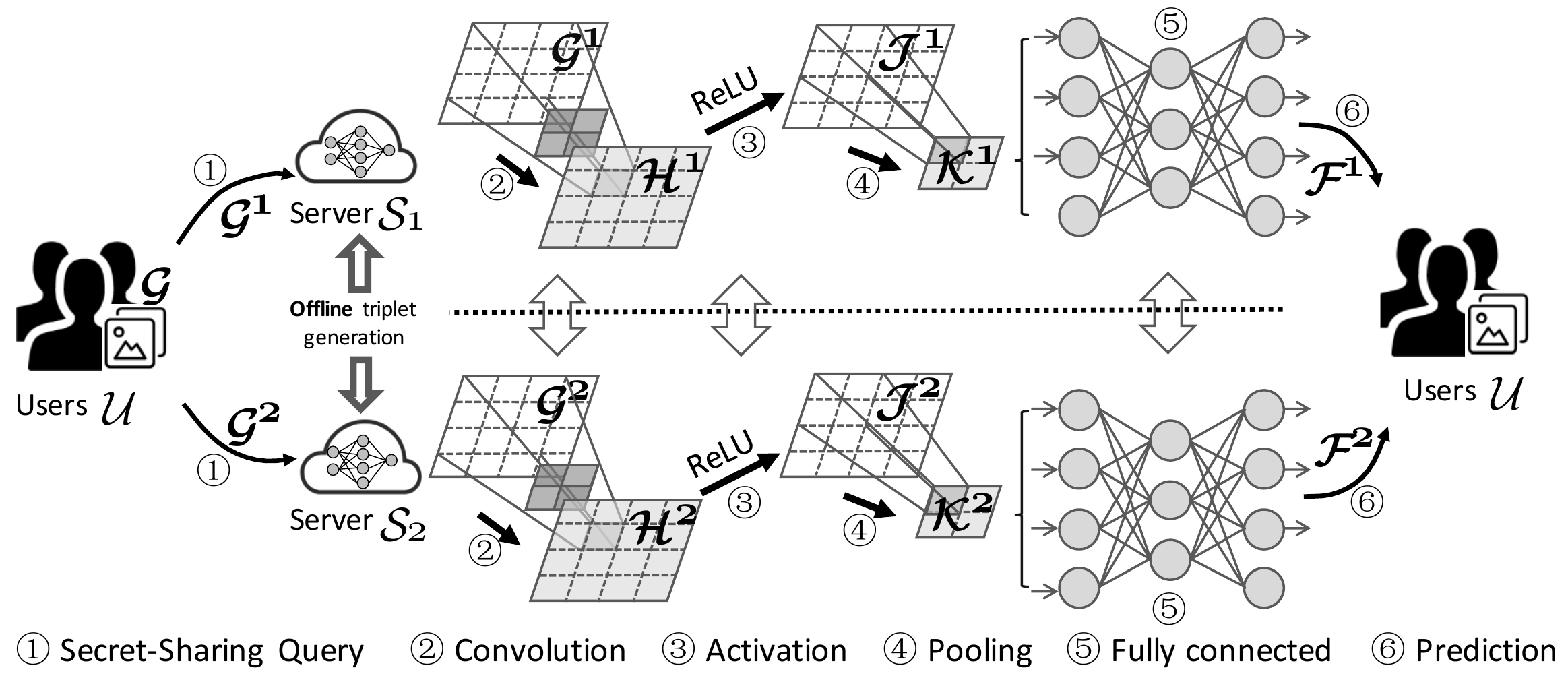}
\caption{An overview of our scheme: ReLU means
rectified linear unit, which is an activation function. 
The notations $\mathcal{G}, \mathcal{H}, \mathcal{J}, \mathcal{K}, \mathcal{F}$ are described in Table~\ref{tab:notation}. 
The superscript $1$ or $2$ denotes the secret-shared information belong to $\sone$ or $\stwo$.}
\label{fig:system_model}
\end{figure*}

\section{Our Construction}

\subsection{System Model and Threat Model}\label{subsec:system_model}
Fig.~\ref{fig:system_model} illustrates our system model.
In an initialization stage,
the model owner (not depicted) secret-shares the model (including the convolutional kernel and the weight in the fully-connected layer) between two servers, $\sone$ and $\stwo$.
One of them also needs to set up a public/private key pair, to be described in Section~\ref{subsec:initialize}.
Such public key is just for accelerating the triplet generation in an offline preparation stage, but not used in the online computation.
Section~\ref{subsec:triplet_generation} discusses the triplet generation in detail.

To use the prediction service, 
user $\ds$ secret-shares the query into two shares and send them to $\sone$ and $\stwo$, respectively, to be described in Section~\ref{subsec:query_distribution}.
$\sone$ interacts with $\stwo$ to make predictions based on the secret shares.

For each stage of the online prediction (including convolution computation, activation computation, pooling computation, fully-connected layer), we design secure interactive protocols to achieve effective computation.
Our protocols rely on the fact that the input (output) of each stage is secret-shared between two servers, and thus the sum of computation results from the two shares is equal to the computation result from the original input.
Section~\ref{subsec:conv_comp} to Section~\ref{subsec:fully_connected} discuss the computation of each layer.
To further reduce the online overheads, we design the accelerated asynchronous online computations as presented in Section~\ref{subsec:async_comp}.
After interactions, the servers derive and return their corresponding shared results to $\ds$, who can then recover the real prediction result locally.

We assume that $\sone$ and $\stwo$ are honest-but-curious and do not collude with each other.
Most service providers 
are motivated to maintain their reputation instead of risking it with collusion.
With this assumption, our goal becomes ensuring both $\sone$ and $\stwo$ cannot learn any information about the model, the query from~$\ds$, and the (intermediate) prediction results.
Intuitively, this gives us hope for higher efficiency without greatly affecting accuracy.

\subsection{Accelerated Triplet Generation}\label{subsec:triplet_generation}
The convolutional and fully-connected layers involve many inner-product computations.
To reduce the cost of online computation over secret shares.
$\sone$ and $\stwo$ first prepare the multiplicative triplets~\cite{Beaver91a}.
Specifically,
$\sone$ holds $a_1$ and $b_1$, $\stwo$ holds $a_2$ and $b_2$.
They want to compute the multiplication of $a = a_1 + a_2$ and $b = b_1 + b_2$
in a shared form \ie, $\sone$ obtains $z_1$, $\stwo$ obtains $z_2$
where $z_1 + z_2 = ab$.
Fig.~\ref{fig:multiplication} presents the triplets generation algorithm $\mathsf{TRIP}$. 
We have $ab = (a_1+a_2)(b_1+b_2) = a_1b_1 + a_1b_2 + a_2b_1 + a_2b_2$.
It is easy for $\sone$ and $\stwo$ to locally compute $a_1b_1$ and $a_2b_2$, respectively.

To compute $a_1b_2$ and $a_2b_1$,
$\sone$ first encrypts $a_1$ and $b_1$ and sends the ciphertexts to $\stwo$.
So, $\stwo$ can add $V_1$, $V_2$, 
which are HE encryption of 
$a_1b_2$ and $a_2b_1$ respectively, 
and a random number $r$ together to obtain $V_3$ via additive homomorphism, 
without any information leakage.
The shared result of $\stwo$ is $z_2=a_2b_2 - r$.
By decrypting $V_3$, $\sone$ obtains $v$, which is added to $a_1b_1$ to obtain the shared result $z_1 = ab - a_2b_2 + r$.
In this way, the shared results $z_1$ and $z_2$ can recover the multiplication result $z$ by simple addition, but neither $\sone$ nor $\stwo$ can learn~$z$.
This preparation is done in an offline phase.

The triplet generation involves
many homomorphic operations.
To further speed up the triplet generation process, we adopt data packing and asynchronous computation.

\begin{figure}[!t]
\fbox{
\begin{minipage} [t]{0.94 \linewidth}
\noindent
$\mathsf{tri} \leftarrow \textsf{TRIP}(a_1, b_1, (\pk, \sk); a_2, b_2, \pk)$\\
\rule{\textwidth}{0.2mm}
\algorithmicrequire\xspace
$\sone$ holds 
$a_1, b_1\in \mathbb{Z}_{2^t}$ and the key pair $(\pk, \sk)$.

$\stwo$ holds the other shares $a_2 = (a - a_1) \bmod {2^t}$, $b_2 = (b - b_1) \bmod {2^t}$ and the public key $\pk$.\\
\algorithmicensure\xspace
$\sone$ gets $\mathsf{tri}^1 = (a_1, b_1, z_1)$.
$\stwo$ gets $\mathsf{tri}^2 = (a_2, b_2, z_2)$, where $z = z_1 + z_2 =a \cdot b\bmod {2^t}$.

\underline{At $\sone$ side:}
\vspace{-2pt}
\begin{enumerate}
\item Encrypt $a_1$ and $b_1$ to have $\llbracket a_1 \rrbracket_{\pk}$ and $\llbracket b_1 \rrbracket_{\pk}$;
\item Send the ciphertexts $\llbracket a_1 \rrbracket_{\pk}$ and $\llbracket b_1 \rrbracket_{\pk}$ to $\stwo$.
\end{enumerate}

\underline{At $\stwo$ side}:
\vspace{-2pt}
\begin{enumerate}
\item Compute $V_1 = \llbracket a_1 \cdot b_2 \rrbracket_{\pk} = \llbracket a_1 \rrbracket_{\pk} \otimes \llbracket b_2 \rrbracket_{\pk}$,\\
$V_2 = \llbracket a_2 \cdot b_1 \rrbracket_{\pk} = \llbracket b_1 \rrbracket_{\pk} \otimes \llbracket b_1 \rrbracket_{\pk}$;
\item Send $V_3= V_1 \oplus V_2\oplus \llbracket r \rrbracket_{\pk}$ to $\sone$,  where $r \in \mathbb{Z}_{2^t}$;
\item Set $z_2 = (a_2 \cdot b_2 - r) \bmod {2^t}$.
\end{enumerate}

\underline{At $\sone$ side:}
\vspace{-2pt}
\begin{enumerate}
\item Decrypt $V_3$ to have $v = a_1 \cdot b_2 +a_2 \cdot b_1 + r$;
\item Set $z_1 = (v+ a_1 \cdot b_1) = (a \cdot b - a_2 \cdot b_2 + r )\bmod {2^t}$.
\end{enumerate}
The triplet is $\mathsf{tri} = ((a_1, a_2), (b_1, b_2), (z_1, z_2))$.
\end{minipage}
}
\caption{Secure triplet generation protocol \textsf{TRIP}}
\label{fig:multiplication}
\end{figure}

\subsubsection{Data Packing}
Considering $n$ shares $\langle a_1^i, b_1^i, a_2^i, b_2^i \rangle_{i=1}^{n}$, we pack $\langle a_1^i \rangle_{i=1}^{n}$ together to be a single plaintext $A_1$ to generate $n$ triplets simultaneously.
Analogously, we compute the packed data $B_1 = \langle b_1^i \rangle_{i=1}^{n}$, $A_2 = \langle a_2^i \rangle_{i=1}^{n}$, and $B_2 = \langle b_2^i \rangle_{i=1}^{n}$.
The homomorphic operations are then performed on the packed data
via SIMD~\cite{dcc/SmartV14} to get the packed triplets $((A_1, A_2), (B_1, B_2), (Z_1, Z_2)) $.
They can then be unpacked to extract the real triplets $\langle (a_1^i, a_2^i), (b_1^i, b_2^i), (z_1^i, z_2^i) \rangle_{i=1}^{n} $, where $z_1^i + z_2^i = $ $(a_1^i + a_2^i)\cdot(b_1^i + b_2^i)$.
In essence, our approach reduces the encryption, transmission, addition, and multiplication costs.
We refer to Appendix~\ref{app:figure} for a toy example for the packed triplet generation process.

\subsubsection{Asynchronous Computation}
$\sone$ and $\stwo$ have to wait for the intermediate results from the other while generating the triplets.
We call this synchronous computation.
To speed it up, we design an asynchronous computation scheme. 
Instead of waiting for the feedback, the servers continue the remaining operations that do not involve the feedback.
For example, $\sone$ can encrypt $b_1$ when transforming the ciphertext $a_1$.
$\stwo$ can encrypt the random number $r$ and compute $a_2 b_2$ ahead of time.
Such asynchronous computation reduces the waiting time and the latency.
See Appendix~\ref{app:figure} for an illustration.

\subsection{Our Scheme}

\subsubsection{Initialization}\label{subsec:initialize}
$\sone$ possesses one share of the prediction model, \ie, the convolutional kernel $\mathsf{C}^1$ and the weight matrix~$\mathsf{W}^1$, and the HE key pair $(\pk, \sk)$. 
$\stwo$ owns the other share of the prediction model (\ie, $\mathsf{C}^2$ and $\mathsf{W}^2$) and the public key $\pk$.
$\ds$ holds the query image $\mathcal{G}$.
They then engage in
the
$\textsf{TRIP}$
protocol
to generate enough number of multiplicative triplets 
$\mathsf{tri} = ((A_1, A_2), (B_1, B_2), (Z_1, Z_2))$.

\subsubsection{Secret-Sharing the Query}\label{subsec:query_distribution}
$\ds$ randomly secret-shares the pixel $\mathcal{G}_{i,j}$ ($i, j \in [1, n]$) of query $\mathcal{G}$ into $\mathcal{G}^1_{i,j}$ and $\mathcal{G}^2_{i,j}$ as $\mathcal{G}_{i,j}= \mathcal{G}^1_{i,j}+\mathcal{G}^2_{i,j} \bmod{2^t}$, 
where $\mathcal{G}^1_{i,j}$ is chosen from $[0, 2^t-1]$ uniformly at random.
$\mathcal{G}^1$ and $\mathcal{G}^2$ are distributed to $\sone$ and $\stwo$, respectively.
From the perspective of the server, the received query, either as $\mathcal{G}^1$ or $\mathcal{G}^2$, 
is random by itself.

\begin{figure}[!t]
\fbox{
\begin{minipage} [t]{0.94 \linewidth}

\noindent $(\mathcal{H}^1; \mathcal{H}^2)\leftarrow \textsf{CONV}(\mathcal{G}^1, \mathsf{C}^1, (\pk, \sk); \mathcal{G}^2, \mathsf{C}^2, \pk)$

\rule{\textwidth}{0.2mm}
\algorithmicrequire\xspace
$\sone$ holds the shares of image $\mathcal{G}^1$, the convolutional kernel $\mathsf{C}^1$, and the shared triplet ($A_1=\langle a_1^k\rangle$, $B_1=\langle b_1^k\rangle$, $Z_1=\langle z_1^k\rangle$);

$\stwo$ holds shares $\mathcal{G}^2$, $\mathsf{C}^2$, and shared triplet ($A_2=\langle a_2^k\rangle$, $B_2=\langle b_2^k\rangle$, $Z_2=\langle z_2^k\rangle$).

We have $z_1^k + z_2^k= (a_1^k + a_2^k) (b_1^k + b_2^k)$ or $Z_1 + Z_2 = AB$.
\\
\algorithmicensure\xspace
$\sone, \stwo$ get shared convoluted images $\mathcal{H}^1, \mathcal{H}^2$ respectively, 
where $\mathcal{H}^1 + \mathcal{H}^2 = (\mathcal{G}^1+\mathcal{G}^2) \cdot (\mathsf{C}^1 + \mathsf{C}^2)$.\\
\underline{$\sone$ and $\stwo$:}
\vspace{-2pt}
\begin{enumerate}
\item \textbf{for} $i, j$ \textbf{in} range $n$
\item[-] Choose the sub-image $\langle \mathcal{G}^1_{p,q} \rangle_{p, q= i - \frac{m-1}{1}}^{p,q =i + \frac{m-1}{1}} $
    as vector $\tilde{\mathcal{G}}^1_{i,j}$
    and $\langle \mathcal{G}^2_{p,q} \rangle_{p, q= i - \frac{m-1}{1}}^{p,q = i + \frac{m-1}{1}} $
    as vector $\tilde{\mathcal{G}}^2_{i,j}$;
\item[-] $\sone$ sends $U^1 = \tilde{\mathcal{G}}^1_{i,j} - A_1, V^1 = \mathsf{C}^1 - B_1$ to $\stwo$,\\
    $\stwo$ sends $U^2 =\tilde{\mathcal{G}}^2_{i,j} - A_2, V^2 = \mathsf{C}^2 - B_2$ to $\sone$;
\item[-] $\sone$ and $\stwo$ compute $U = \tilde{\mathcal{G}}_{i,j} - A, V = \mathsf{C} - B$;
\item[-] $\sone$ computes $\mathcal{H}^1_{i,j} = -UV + \tilde{\mathcal{G}}^1_{i,j}V + \mathsf{C}^1U + Z_1$,\\
    $\stwo$ computes $\mathcal{H}^2_{i,j} = \tilde{\mathcal{G}}^2_{i,j}V + \mathsf{C}^2U + Z_2$.
\item Return
$\mathcal{H}^1$ and $\mathcal{H}^2$.
\end{enumerate}
\end{minipage}
}
\caption{Secure convolution computation protocol \textsf{CONV}}
\label{fig:convolutional}
\end{figure}

\subsubsection{Convolution Computation}\label{subsec:conv_comp}
Fig.~\ref{fig:convolutional} describes our secure convolution computation protocol ($\textsf{CONV}$).
Given the secret-shared query ($\mathcal{G}^1$, $\mathcal{G}^2$) and 
the secret-shared convolutional kernel $(\mathsf{C}^1$, $\mathsf{C}^2$) as input, 
the two servers run the $\textsf{CONV}$ protocol and output the shared results $\mathcal{H}^1$ and $\mathcal{H}^2$  
such that $\mathcal{H}^1 + \mathcal{H}^2 = \mathcal{G}\cdot \mathsf{C}$.

In more detail, 
$\sone$ and $\stwo$ apply convolution over each sub-image $\tilde{\mathcal{G}}^1_{i,j}$ and $\tilde{\mathcal{G}}^2_{i,j}$ of the size $m\times m$ from the shared images $\mathcal{G}^1$ and $\mathcal{G}^2$, respectively.
With $\mathsf{tri}$ prepared by the execution of $\textsf{TRIP}$ in the offline phase,
$\sone$ ($\stwo$) uses $A_1$ ($A_2$) as a one-time pad to hide the sub-images $\tilde{\mathcal{G}}^1_{i,j}$ (resp. $\tilde{\mathcal{G}}^2_{i,j}$)).
Likewise,
they use $B_1$ ($B_2$) to hide the convolutional kernel $\mathsf{C}^1$ (resp. $\mathsf{C}^2$).
After they exchanged these padded shares, they can locally compute $\mathcal{H}^1$ and $\mathcal{H}^2$.
For correctness:

\begin{align*}
&\mathcal{H}^1 + \mathcal{H}^2\\
=& (-UV + \tilde{\mathcal{G}}^1_{i,j}V + \mathsf{C}^1U + Z_1) + (\tilde{\mathcal{G}}^2_{i,j}V + \mathsf{C}^2U + Z_2)\\
=& -UV + (\tilde{\mathcal{G}}^1_{i,j}V + \tilde{\mathcal{G}}^2_{i,j}V) + (\mathsf{C}^1U + \mathsf{C}^2U) + (Z_1 + Z_2)\\
=& - (\tilde{\mathcal{G}}_{i,j} -A)V + \tilde{\mathcal{G}}_{i,j}V + \mathsf{C}U + AB\\
=& AV + \mathsf{C}U + AB\\
=& A(\mathsf{C} - B) + \mathsf{C}(\tilde{\mathcal{G}}_{i,j} - A) + AB\\
=& \tilde{\mathcal{G}}_{i,j} \cdot \mathsf{C}.
\end{align*}

\subsubsection{Activation Computation}\label{subsec:activation_comp}

\begin{figure}[!t]
\centering
\fbox{\parbox{0.39\textwidth}{%
\includegraphics[width=0.8\columnwidth]{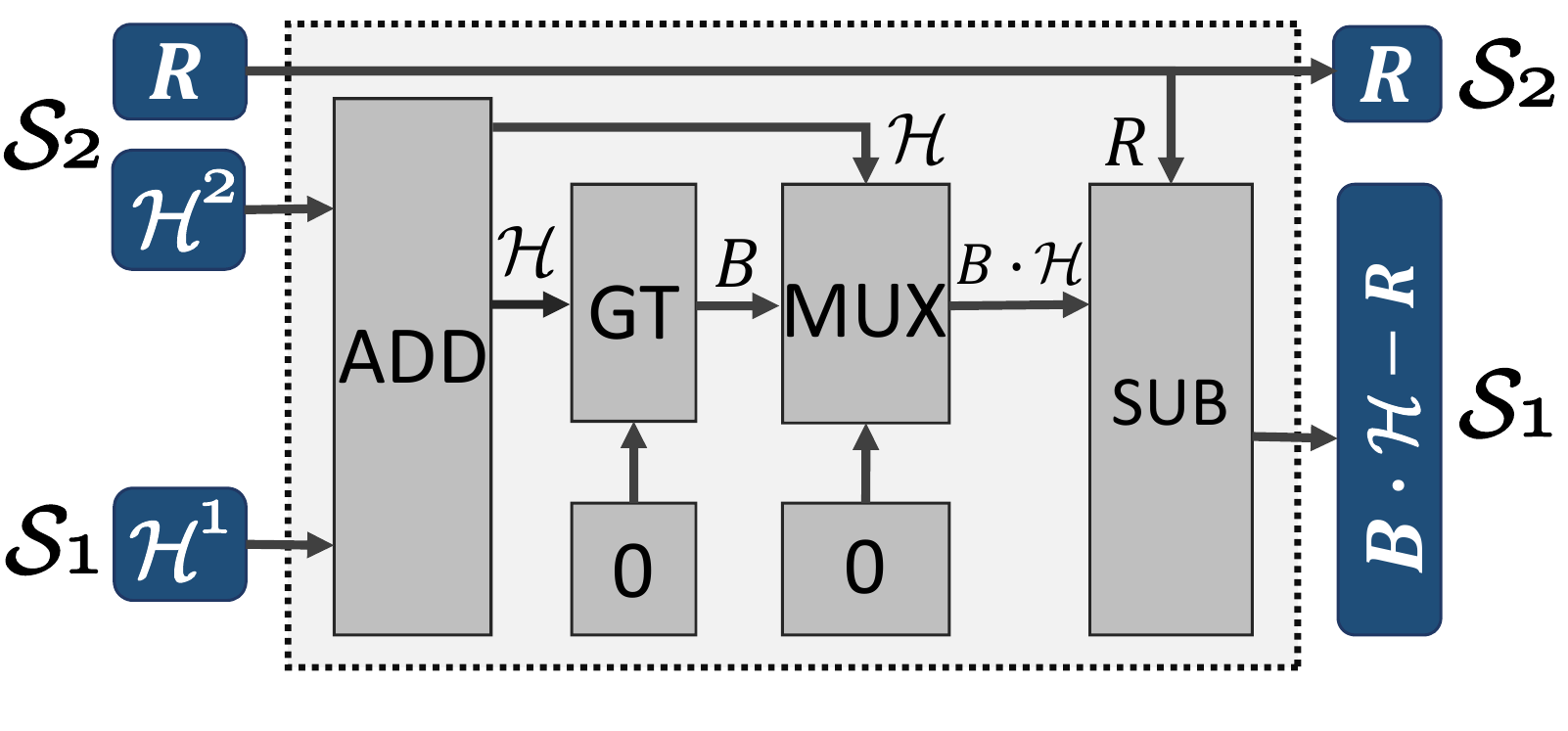}
}}
\caption{Our (garbled) circuits for ReLU activation function $\mathsf{ActF}$}
\label{fig:activation_function}
\end{figure}

$\sone$ and $\stwo$ use the garbled circuit $\mathsf{ActF}$ in Fig.~\ref{fig:activation_function} to compute the activation function over 
$\mathcal{H}^1$ and $\mathcal{H}^2$.
The circuit utilizes four sub-circuits:
\begin{itemize}[leftmargin=10pt]
\item $\mathsf{ADD}(a, b)$ outputs the sum $a+b$;
\item $\mathsf{GT}(a, b)$ returns the bit denoting if $a>b$;
\item $\mathsf{MUX} (a, b, c)$ outputs $a$ or $b$ according to $c$,
\ie, if $c$ holds, $\mathsf{MUX} (a, b, c) = b$, otherwise $\mathsf{MUX} (a, b, c)=a$; and
\item $\mathsf{SUB}(a, b)$ returns the difference $a-b$.
\end{itemize}

$\sone$ and $\stwo$ run $\mathsf{ActF}$ with $\mathcal{H}^1$ and $(\mathcal{H}^2, R)$ being the respective inputs, where $R$ is a random matrix generated by $\stwo$.
The outputs of $\sone$ and $\stwo$ are $(0|1)\cdot\mathcal{H}-R$ and $R$, respectively, 
where $(0|1)$ denotes a variable of value either $0$ or $1$ depending on $\mathcal{H}$.
To summarize in the standard S2C notation, 
$((0|1)\cdot\mathcal{H}-R; R) \leftarrow \mathsf{ActF}(\mathcal{H}^1; (\mathcal{H}^2, R))$.

Specifically, $\mathsf{ADD}(\mathcal{H}^1, \mathcal{H}^2)$
adds $\mathcal{H}^1$ and $\mathcal{H}^2$ to get $\mathcal{H}$.
Then, $\mathsf{GT}(\mathcal{H}, 0)$ compares $\mathcal{H}$ and~$0$ element-wise for ReLU.
The output is a binary matrix $B$ of the same size as $\mathcal{H}$, where the pixel $B_{i,j} = 1$ if $\mathcal{H}_{i,j} > 0$, $0$ otherwise.
With $B$, $\mathsf{MUL}(\mathcal{H},0, B)$ performs activation function,
\ie, if $\mathcal{H}_{i,j}>0$ ($B_{i,j}=1$), outputs $\mathcal{H}$, otherwise outputs $0$.
Finally, $\mathsf{SUB}$ makes element-wise subtraction to have $B\cdot\mathcal{H}-R$,
which is the output of $\sone$.
Let $\mathcal{J}$ be the activated image $B\cdot\mathcal{H}$.
$\mathcal{J}^1=B\cdot\mathcal{H}-R$.
The output of $\stwo$ is regarded as $\mathcal{J}^2=R$.

\subsubsection{Pooling Computation}\label{subsec:pooling_comp}

Fig.~\ref{fig:pooling} shows the $\mathsf{POOL}$ ``protocol'' for element-wise average-pooling over $(\mathcal{J}^1, \mathcal{J}^2)$ to obtain $(\mathcal{K}^1, \mathcal{K}^2)$.
$\sone$ and $\stwo$ run $\mathsf{POOL}$ with the activated shared data $\mathcal{J}^1$ and $\mathcal{J}^2$ being the respective inputs. 
The outputs of $\sone$ and $\stwo$ are the secret-shared pooled results $\mathcal{K}^1$ and $\mathcal{K}^2$, respectively.

The pixel value $\mathcal{K}_{i,j}^1$ is the average of the corresponding $q \times q$ pixels in $\mathcal{J}^1$.
Analogously, $\mathcal{K}_{i,j}^2$ can be derived from $\mathcal{J}^2$.
Employing the average value to replace the original pixels, we can reduce $\mathcal{J}$ to the down-sampled image $\mathcal{K}$ of size $\lceil \frac{n}{q}\rceil \times \lceil \frac{n}{q} \rceil$.
\begin{align*}
&{(\mathcal{J}_1^1+\mathcal{J}_2^1+ \cdots+\mathcal{J}_{q^2}^1)} / {q^2} + {(\mathcal{J}_1^2+\mathcal{J}_2^2+ \cdots+\mathcal{J}_{q^2}^2)} / {q^2}\\
=&{(\mathcal{J}_1+\mathcal{J}_2+ \cdots+ \mathcal{J}_{q^2})} / {q^2}.
\end{align*}

It is easy to see that they can perform the above computation locally without any interaction.

\begin{figure}[!t]
\fbox{
\begin{minipage} [t]{0.94\linewidth}
\noindent $(\mathcal{K}^1; \mathcal{K}^2)\leftarrow \mathsf{POOL}(\mathcal{J}^1; \mathcal{J}^2)$

\rule{\textwidth}{0.2mm}
\algorithmicrequire\xspace
$\sone$ holds the shared activated image $\mathcal{J}_1$;

$\stwo$ holds the other shares $\mathcal{J}_2$.\\
\algorithmicensure\xspace
$\sone$ obtains the shared pooled image $\mathcal{K}_1$;

$\stwo$ holds the other share $\mathcal{K}_2$. $\mathcal{K}_1 + \mathcal{K}_2 = pooling (\mathcal{J}_1 + \mathcal{J}_2)$.
\\
\underline{At $\sone$ side:}
\vspace{-2pt}
\begin{enumerate}
\item \textbf{for} $i, j$ \textbf{in} range $\lfloor \frac{n}{q} \rfloor$
\item[-] Compute $\frac{1}{q \times q}$ $\sum_{u=0}^{u=q-1}\sum_{v=0}^{v=q-1}\mathcal{J}^1_{qi-u, qj-v}$.
\item[-] Set the averages as the value of $\mathcal{K}^1_{i,j}$.
\end{enumerate}
\underline{At $\stwo$ side}:
\vspace{-2pt}
\begin{enumerate}
\item \textbf{for} $i, j$ \textbf{in} range $\lfloor \frac{n}{q} \rfloor$
\item[-] Compute $\frac{1}{q \times q}$ $\sum_{u=0}^{u=q-1}\sum_{v=0}^{v=q-1}\mathcal{J}^2_{qi-u, qj-v}$.
\item[-] Set the averages as the value of $\mathcal{K}^2_{i,j}$.
\end{enumerate}

\end{minipage}
}
\caption{Secure pooling computation protocol}
\label{fig:pooling}
\end{figure}

\begin{figure}[!ht]
\centering
\includegraphics[width=0.9\columnwidth]{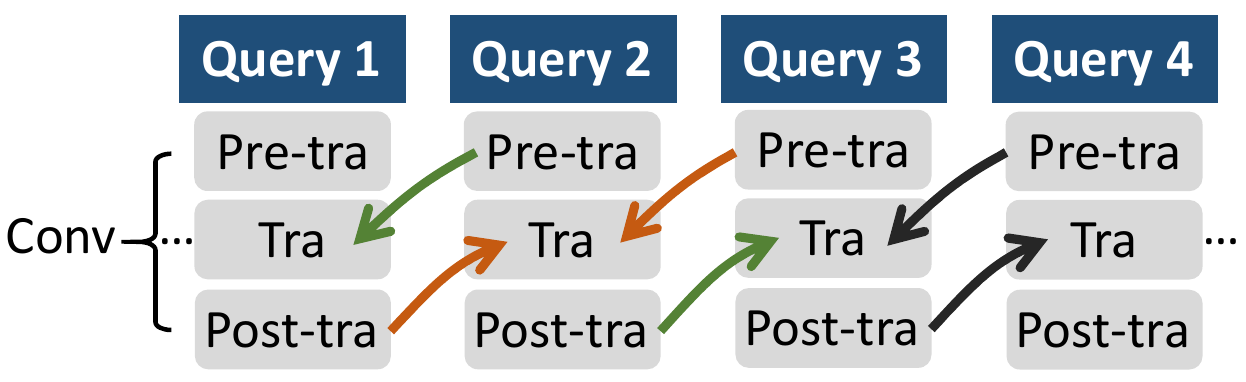}
\caption{Asynchronous computation for online prediction}
\label{fig:online_async_comp}
\end{figure}

\subsubsection{Fully-Connected-Layer Computation}\label{subsec:fully_connected}
Finally, the fully-connected layer, in essence, performs dot products between the pooled data $\mathcal{K}$ and the weight parameters $\mathsf{W}$, which can be computed using the triplets
similar to that in the convolutional layer illustrated in Fig.~\ref{fig:convolutional}.
We skip the largely repetitive details.
Specifically, $\sone$ and $\stwo$ take as input the shares $\mathcal{K}^1, \mathsf{W}^1$
and $\mathcal{K}^2, \mathsf{W}^2$ respectively,
resulting in the shared prediction results $\mathcal{F}^1$ and $\mathcal{F}^2$.
User $\ds$ can merge
$\mathcal{F}^1$ and $\mathcal{F}^2$ to recover the
prediction
result $\mathcal{F} = \mathcal{K}\mathsf{W}$.

\subsection{Accelerated Asynchronous Online Computations}\label{subsec:async_comp}
Finally, we remark that we can accelerate convolution and fully-connected layers by 
simultaneously conducting computations (of $U^1, V^1, U^2, V^2$ (pre-transmission), $U, V, \mathcal{H}^1_{i,j}, \mathcal{H}^2_{i,j}$ (post-transmission))
and transmission (of $U^1, V^1, U^2, V^2$).
Fig.~\ref{fig:online_async_comp} depicts asynchronous computations in the convolution layers.
The computation process at the tail of the arrow is in parallel with the transmission process at the arrowhead.
For example, the pre-transmission and post-transmission operations of query 2
are in parallel with the transmission of query 1 and query 3, respectively.

\section{Performance Evaluation}
\subsection{Experimental Setup and Dataset}\label{subsec:exp}
Our experiments use separate machines for $\ds$, $\sone$, and $\stwo$.
Each has an Intel $4$-Core CPU operating at $1.60$ GHz with $8$ GB RAM, running Ubuntu 18.04 as the operating system.
We implemented our scheme in C++ with Python binding.
For garbled circuits, we use the ABY library with SIMD circuits~\cite{ndss/Demmler0Z15}.
Following MiniONN, we use YASHE for FHE and implement it by the SEAL library~\cite{pieee/DowlinGLLNW17}, supporting SIMD.
In YASHE, the degree of the polynomial $n$ is set to be $4096$. In this way, we can pack $4096$ elements together.
The plaintext modulus is $101,285,036,033$. The length of the plaintext is $64$-bit.
The ciphertext modulus is $128$-bit.
These parameters matter in Section~\ref{subsec:eff_pack}.
All results are averaged over at least $5$ runs, in which the error is controlled within $3\%$.

We conduct experiments over two standard datasets, MNIST~\cite{lecun1998mnist} and CIFAR-10~\cite{KrizhevskyH09}.
The MNIST dataset~\cite{lecun1998mnist} consists of $60,000$ grayscale images of hand-written digit belong to $10$ classes, each $28\times 28$ pixels. 
The CIFAR-10 dataset~\cite{KrizhevskyH09} contains $60,000$ color images of size $32\times 32$ pixels in $10$ different classes.

Apart from real datasets, a realistic neural network model is also important to demonstrate our system performance.
Fig.~\ref{fig:neural_network_architecture_on_MNIST} and Fig.~\ref{fig:neural_network_architecture_on_CIFAR10} detail the neural network architectures we used 
(following MiniONN~\cite{LiuJLA17}) on MNIST and CIFAR-10 dataset, respectively.

\begin{figure}[!t]
\fbox{
\begin{minipage} [t]{0.94\linewidth}

\begin{enumerate}
\item \emph{Convolution}: input image $28\times 28$, window size $5\times 5$, stride$ (1,1)$, number of output channels of $16$: $\mathbb{R}^{16\times 576}\leftarrow \mathbb{R}^{16\times 25} \times \mathbb{R}^{25\times 576}$.
\item \emph{ReLU Activation}: calculates ReLU for each input.
\item \emph{Average Pooling}: window size $1\times 2\times 2$ and outputs $\mathbb{R}^{16\times 12\times 12}$.
\item \emph{Convolution}: input image $12\times 12$, window size $5\times 5$, stride $(1,1)$, number of output channels of $16$: $\mathbb{R}^{16\times 64}\leftarrow \mathbb{R}^{16\times 400} \times \mathbb{R}^{400 \times 64}$.
\item \emph{ReLU Activation}: calculates ReLU for each input.
\item \emph{Average Pooling}: window size $1\times 2\times 2$ and outputs $\mathbb{R}^{16\times 4\times 4}$.
\item \emph{Fully Connected}: fully connects the incoming $256$ notes to the outgoing $100$ nodes: $\mathbb{R}^{100\times 1} \leftarrow \mathbb{R}^{100\times 256} \times \mathbb{R}^{256\times 1}$.
\item \emph{ReLU Activation}: calculates ReLU for each input.
\item \emph{Fully Connected}: fully connects the incoming $100$ notes to the outgoing $10$ nodes: $\mathbb{R}^{10\times 1} \leftarrow \mathbb{R}^{10\times 100} \times \mathbb{R}^{100\times 1}$.
\end{enumerate}

\end{minipage}
}
\caption{The neural network architecture on MNIST dataset}
\label{fig:neural_network_architecture_on_MNIST}
\end{figure}

\begin{figure}[!t]
\fbox{
\begin{minipage} [t]{0.94\linewidth}
\begin{enumerate}
\item \emph{Convolution}: input image $32\times 32$, window size $3\times 3$, stride$ (1,1)$, number of output channels of $64$: $\mathbb{R}^{64\times 1024}\leftarrow \mathbb{R}^{64\times 27} \times \mathbb{R}^{27\times 1024}$.
\item \emph{ReLU Activation}: calculates ReLU for each input.
\item \emph{Convolution}: window size $3\times 3$, stride$ (1,1)$, number of output channels of $64$: $\mathbb{R}^{64\times 1024}\leftarrow \mathbb{R}^{64\times 576} \times \mathbb{R}^{576\times 1024}$.
\item \emph{ReLU Activation}: calculates ReLU for each input.
\item \emph{Average Pooling}: window size $1\times 2\times 2$ and outputs $\mathbb{R}^{64\times 16\times 16}$.
\item \emph{Convolution}: window size $3\times 3$, stride $(1,1)$, number of output channels of $64$: $\mathbb{R}^{64\times 256}\leftarrow \mathbb{R}^{64\times 576} \times \mathbb{R}^{576 \times 256}$.
\item \emph{ReLU Activation}: calculates ReLU for each input.
\item \emph{Convolution}: window size $3\times 3$, stride $(1,1)$, number of output channels of $64$: $\mathbb{R}^{64\times 256}\leftarrow \mathbb{R}^{64\times 576} \times \mathbb{R}^{576 \times 256}$.
\item \emph{ReLU Activation}: calculates ReLU for each input.
\item \emph{Average Pooling}: window size $1\times 2\times 2$ and outputs $\mathbb{R}^{64\times 8\times 8}$.
\item \emph{Convolution}: window size $3\times 3$, stride $(1,1)$, number of output channels of $64$: $\mathbb{R}^{64\times 64}\leftarrow \mathbb{R}^{64\times 576} \times \mathbb{R}^{576 \times 64}$.
\item \emph{ReLU Activation}: calculates ReLU for each input.
\item \emph{Convolution}: window size $1\times 1$, stride $(1,1)$, number of output channels of $64$: $\mathbb{R}^{64\times 64}\leftarrow \mathbb{R}^{64\times 64} \times \mathbb{R}^{64 \times 64}$.
\item \emph{ReLU Activation}: calculates ReLU for each input.
\item \emph{Convolution}: window size $1\times 1$, stride $(1,1)$, number of output channels of $64$: $\mathbb{R}^{16\times 64}\leftarrow \mathbb{R}^{16\times 64} \times \mathbb{R}^{64 \times 64}$.
\item \emph{ReLU Activation}: calculates ReLU for each input.
\item \emph{Fully Connected}: fully connects the incoming $1024$ notes to the outgoing $10$ nodes: $\mathbb{R}^{10\times 1} \leftarrow \mathbb{R}^{10\times 1024} \times \mathbb{R}^{1024\times 1}$.
\end{enumerate}
\end{minipage}
}
\caption{The neural network architecture on the CIFAR-10 dataset}
\label{fig:neural_network_architecture_on_CIFAR10}
\end{figure}

\begin{figure*}[!t]
\centering
\begin{minipage}[!htbp]{0.48\linewidth}
\centering
\includegraphics[width=0.912\textwidth]{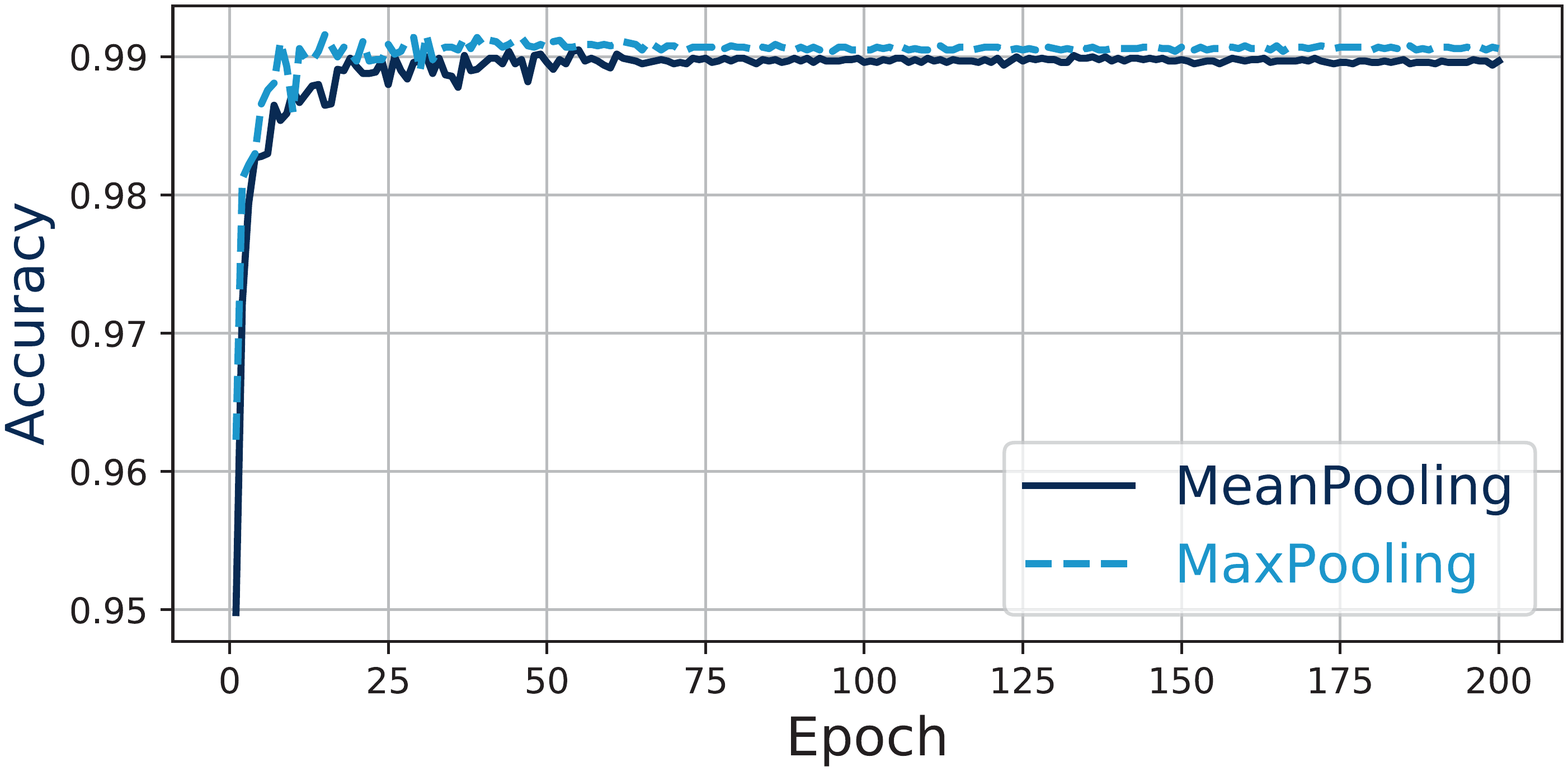}
\caption{ Accuracy of the trained model with different pooling methods on MNIST dataset}
\label{fig:mnionn_acc_MeanPooling}
\end{minipage}%
\hspace{10pt}
\begin{minipage}[!htbp]{0.48\linewidth}
\centering
\includegraphics[width=0.912\textwidth]{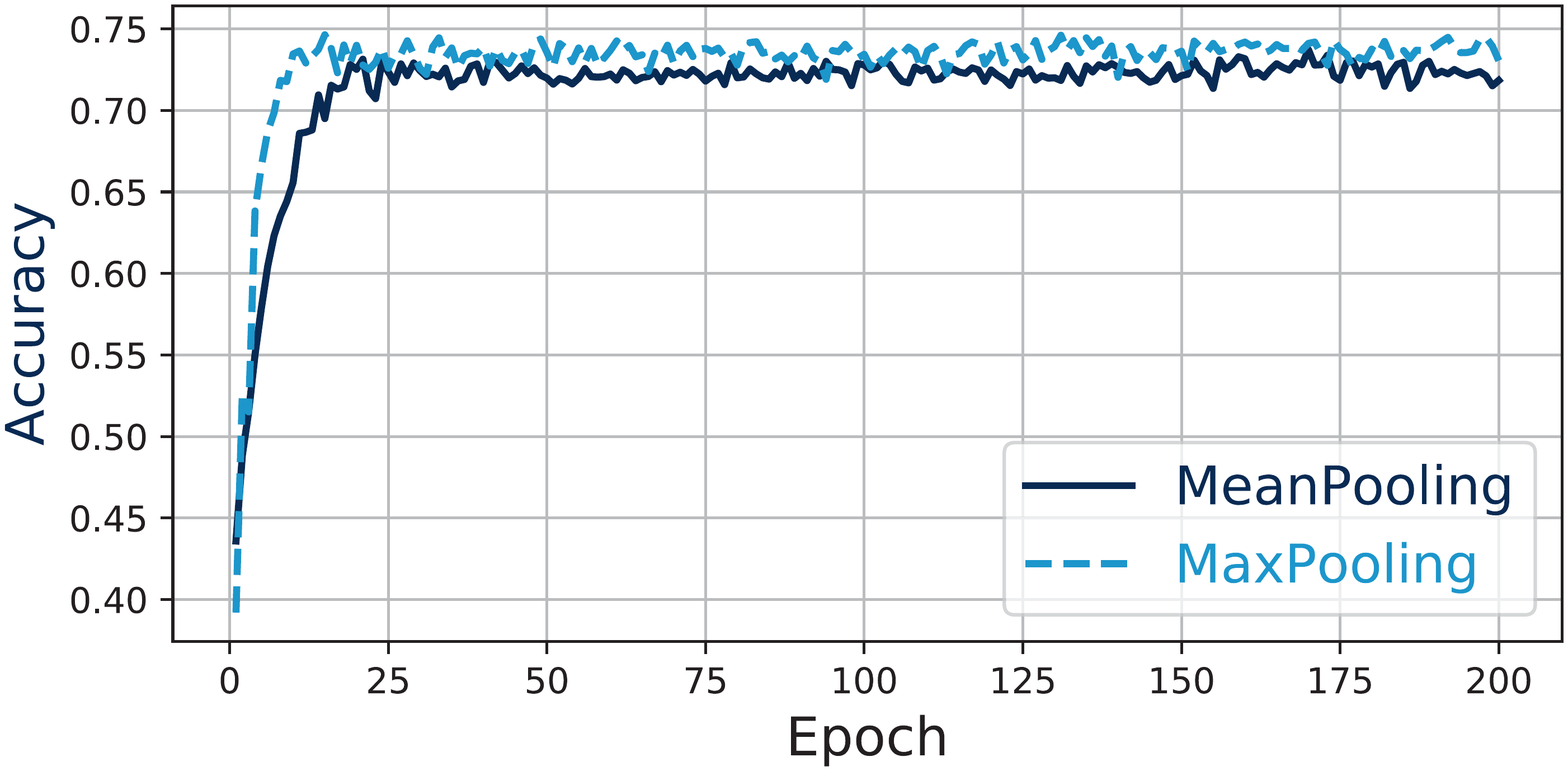}
\caption{Accuracy of the trained model with different pooling methods on CIFAR-10 dataset}
\label{fig:mnionn_acc_MaxPooling}
\end{minipage}
\end{figure*}

\begin{figure}[!t]
\centering
\includegraphics[width=0.4\textwidth]{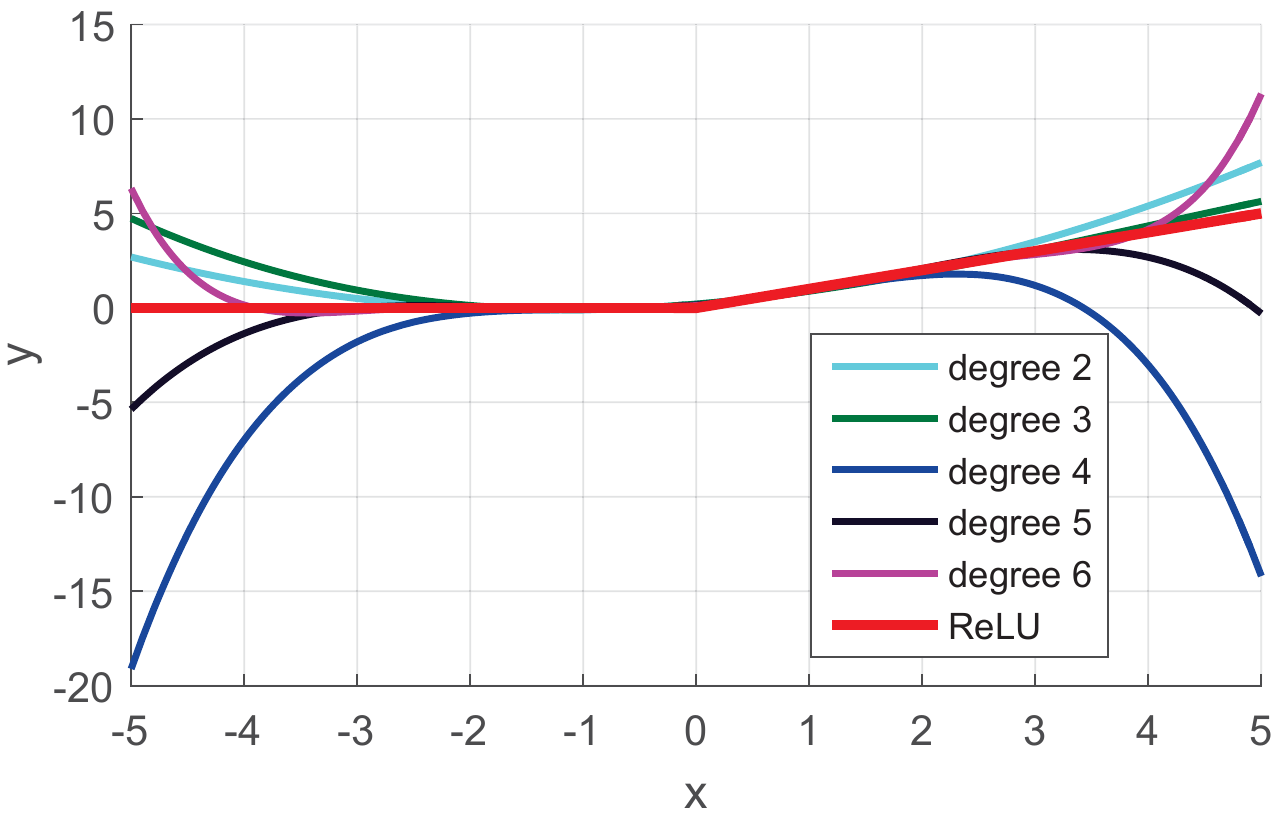}
\caption{Approximate function of ReLU activation function}
\label{fig:ReluPoly}
\end{figure}

\subsection{Accuracy Evaluation}
\subsubsection{Effect of Pooling Method}

To evaluate the effect of average-pooling and max-pooling, we trained the model on the MNIST and CIFAR-10 datasets for each pooling method.
Fig.~\ref{fig:mnionn_acc_MeanPooling} and Fig.~\ref{fig:mnionn_acc_MaxPooling} plot the accuracy against the training epoch for the MNIST and CIFAR datasets, respectively.
The light blue line represents the accuracy using max-pooling, and the dark blue line shows the accuracy using average-pooling.

The accuracy grows with the epochs until stabilized.
For the MNIST dataset, after $10$ epochs, the accuracy of max-pooling reaches $98.7\%$, while that of average-pooling is just $98.2\%$.
After $100$ epochs, the accuracy of max-pooling achieves $99.1\%$, and that of average-pooling achieves $99.0\%$, with a difference of $0.1\%$.
For the CIFAR-10 dataset, after $10$ epochs, max-pooling achieves an accuracy of $72\%$ while average-pooling achieves an accuracy of $65\%$. 
After $100$ epochs, the accuracy of max-pooling is $74\%$, and that of average-pooling is $73\%$, with a difference of $1\%$ again.
To conclude, their accuracy differs by just $1\%$, which is acceptable in most application scenarios.
That said, we remark that it could be too big in life-critical applications such as disease diagnosis. 

\subsubsection{Effect of Activation Function}
Some FHE-based schemes~\cite{Gilad-BachrachD16, HesamifardTGW18} approximate the activation function by polynomials.
Fig.~\ref{fig:ReluPoly} plots the curves of the original ReLU function and its approximations with different degrees,
which are polynomial regression function \emph{polyfit} from the Python package \texttt{numpy}.

The approximation is satisfactory
when the absolute value of the input is smaller than $4$.
However, for larger input, the error introduced by the approximation becomes very large,
which in turn affects the prediction accuracy.
Our scheme
reaches the same accuracy as plaintext computation
by directly designing
the garbled circuits for ReLU, which compute an identical output as ReLU.

\begin{table}[!t]
\setlength{\belowcaptionskip}{3pt}
\centering
\newcolumntype{M}[1]{>{\centering\arraybackslash}m{#1}}
\renewcommand{\arraystretch}{1.6}
\addtolength{\tabcolsep}{-0pt}
\caption{Comparison of triplet generation}
\begin{tabular}{ M{7em}||M{5em}|M{13em} }
\Xhline{1.1pt}
Methodology	&	Complexity	&	Required operations\\
\Xhline{1.1pt}
SecureML~\cite{MohasselZ17}	&	$\mathcal{O}(n)$	&	$n\cdot$($5$Enc + $2$CMul + $2$Add + $1$Dec)\\
\hline
MiniONN~\cite{LiuJLA17}	&	$\mathcal{O}(n/l)$	&	$\frac{n}{l} \cdot$($5$Enc + $2$CMul + $2$Add + $1$Dec)\\
\hline
Our Scheme	&	$\mathcal{O}(n/l)$	&	$\frac{n}{l} \cdot$($5$Enc + $2$CMul + $2$Add + $1$Dec)\\
\Xhline{1.1pt}
\end{tabular}
\begin{tablenotes}
\item [*] Enc: encryption; CMul: constant multiplication over ciphertext; Add: addition between two ciphertexts; Dec: decryption
\end{tablenotes}
\label{tab:efficiency_analysis}
\end{table}

\subsection{Efficiency Evaluation}\label{subsec:eff_pack}
\subsubsection{Triplet Generation}
Table~\ref{tab:efficiency_analysis} compares our triplets generation method with prior work in terms of the computational complexity.
As Fig.~\ref{fig:multiplication}, triplet generation for two $n$-dimensional shared vectors of SecureML~\cite{MohasselZ17} encrypts each element of the shared vector, respectively.
This process consists of $5n$ Enc
encryptions, $2n$ CMul multiplications and additions, $n$
decryptions, in which $n$ ciphertexts are transferred between the two servers.

In contrast, MiniONN~\cite{LiuJLA17} and our scheme use the packing technique.
All the $l$ elements can be computed at the same time.
The $n$-dimensional vector is compressed into an %
$(n/l)$-dimensional vector.
This means that the complexity of MiniONN and our scheme is just
$\mathcal{O}(n/l)$.

Table~\ref{tab:triplet} shows that our packed triplet generation is several order-of-magnitude faster.
Further adopting asynchronous computations improves the efficiency by $13.6\%$, resulting in an overall speedup of $4697\times$.

\begin{table}[!t]
\setlength{\belowcaptionskip}{3pt}
\centering
\newcolumntype{M}[1]{>{\centering\arraybackslash}m{#1}}
\renewcommand{\arraystretch}{1.6}
\addtolength{\tabcolsep}{-0pt}
\caption{Triplet generation costs (ms)}
\begin{tabular}{ M{6em}|M{6em} |M{9em}|M{4.4em} }
 \Xhline{1.1pt}
Original triplet generation	&	Packed triplet \qquad generation	&	Packed triplet generation \qquad with Asyn. Comp.	&	Performance Gain \\
\Xhline{1.1pt}
 $79716.524$	&	$19.635$	&	$16.970$	&	$4697 \times$\\
\Xhline{1.1pt}
\end{tabular}
\label{tab:triplet}
\end{table}

\begin{table}[!t]
\setlength{\belowcaptionskip}{3pt}
\centering
\newcolumntype{M}[1]{>{\centering\arraybackslash}m{#1}}
\renewcommand{\arraystretch}{1.6}
\addtolength{\tabcolsep}{-0pt}
\caption{ReLU costs (ms)}
\begin{tabular}{ M{3em}||M{4.5em}|M{4.5em} |M{5em}|M{5em} }
 \Xhline{1.1pt}
	&	Offline	&	Online	&	Avg. Offline	&	Avg. Online \\
\Xhline{1.1pt}
$\sone$	&	$87.815$	&	$752.545$	&	$0.021$	&	$0.184$\\
\hline
$\stwo$	&	$100.143$	&	$516.413$	&	$0.024$	&	$0.126$\\
\Xhline{1.1pt}
\end{tabular}
\label{tab:relucost}
\end{table}

\begin{table}[!t]
\setlength{\belowcaptionskip}{3pt}
\centering
\newcolumntype{L}[1]{>{\raggedright\arraybackslash}m{#1}}
\newcolumntype{R}[1]{>{\raggedleft\arraybackslash}m{#1}}
\renewcommand{\arraystretch}{1.6}
\addtolength{\tabcolsep}{-0pt}
\caption{Polynomial approximation}
\begin{tabular}{ L{15em}||R{4em}|R{6em} }
 \Xhline{1.1pt}
Approximate Function	&	Time (ms)	&	Performance Gain \\
 \Xhline{1.1pt}
$0.1992 + 0.5002x + 0.1997x^2$	&	$19.02$	&	$61\times$ \\
\hline
$0.1995 + 0.5002x+ 0.1994x^2 - 0.0164x^3$	&	$38.00$	&	$123\times$ \\
\hline
$0.1500 + 0.5012x + 0.2981x^2 - 0.0004x^3 - 0.0388x^4$	&	$69.62$	&	$225\times$ \\
\hline
$0.1488 + 0.4993x + 0.3007x^2+ 0.0003x^3 - 0.0168x^4$	&	$69.64$	&	$224\times$ \\
\hline
$0.1249 + 0.5000x + 0.3729x^2 - 0.0410x^4 + 0.0016x^6$	&	$82.20$	&	$265\times$ \\
\Xhline{1.1pt}
\end{tabular}
\label{tab:poly_relu_comparison}
\end{table}

\begin{table*}[!t]
\setlength{\belowcaptionskip}{3pt}
\centering
\newcolumntype{M}[1]{>{\centering\arraybackslash}m{#1}}
\newcolumntype{R}[1]{>{\raggedleft\arraybackslash}m{#1}}
\renewcommand{\arraystretch}{1.6}
\addtolength{\tabcolsep}{-0pt}
\caption{Performance of each stage on the MNIST and CIFAR-10 datasets}
\begin{tabular}{ M{5em}||M{11em}|R{5em} ||M{11em}|R{5em}|M{11em}|R{5em}}
 \Xhline{1.1pt}
	&	\multicolumn{2}{c||}{MNIST}	&	\multicolumn{4}{c}{CIFAR-10}\\
 \Xhline{1.1pt}
Phases	&	Stages	&	Latency (s)	&	Stages	&	Latency (s)	&	Stages	&	Latency (s)\\
 \Xhline{1.1pt}
 \multirow{2}{*}{Offline}	&	{$\sone$}	&	$0.219$	&	$\sone$	&	$3.711$	&	--	&	--\\
\cline{2-7}
	&	{$\stwo$}	&	$0.252$	&	$\stwo$	&	$4.231$	&	--	&	--\\
\Xhline{0.8pt}
 \multirow{9}{*}{Online}	&	1. Convolution	&	$0.055$	&	1. Convolution	&	$0.425$	&	10. Average Pooling	&	$0.000$\\
\cline{2-7}
	&	2. ReLU Activation	&	$0.425$	&	2. ReLU Activation	&	$20.304$	&	11. Convolution	&	$0.566$\\
\cline{2-7}
	&	3. Average Pooling	&	$0.001$	&	3. Convolution	&	$9.062$	&	12. ReLU Activation	&	$1.268$\\
\cline{2-7}
	&	4. Convolution	&	$0.098$	&	4. ReLU Activation	&	$20.304$	&	13. Convolution	&	$0.062$\\
\cline{2-7}
	&	5. ReLU Activation	&	$0.317$	&	5. Average Pooling	&	$0.001$	&	14. ReLU Activation	&	$1.269$\\
\cline{2-7}
	&	6. Average Pooling	&	$0.000$	&	6. Convolution	&	$2.266$	&	15. Convolution	&	$0.015$\\
\cline{2-7}
	&	7. Fully connected	&	$0.006$	&	7. ReLU Activation	&	$5.075$	&	16. ReLU Activation	&	$0.317$\\
\cline{2-7}
	&	8. ReLU Activation	&	$0.031$	&	8. Convolution	&	$2.265$	&	17. Fully connected	&	$0.002$\\
\cline{2-7}
	&	9. Fully connected	&	$0.001$	&	9. ReLU Activation	&	$5.076$	&	--	&	--\\

 \hline
Total	&	\multicolumn{2}{r||}{$3.835$}	&	\multicolumn{4}{r}{$76.224$}\\
\Xhline{1.1pt}
\end{tabular}
\label{tab:eachphase}
\end{table*}

\subsubsection{Activation Function}
To make our contribution stand out, we run the activation function circuits alone to demonstrate its performance.
Table~\ref{tab:relucost} summarizes our results.
We perform our ReLU with the SIMD circuits and activate all the $4096$ packed data simultaneously.
The offline time costs for circuit generation are $87.815$ms and $100.143$ms, respectively.
The time consumptions in the online phase are $752.545$ms and $516.413$ms.
The average per-data time costs of the offline and online computations are $0.045$ms and $0.310$ms.

Compared with the existing works using approximated polynomials,
our GC-based circuits also provide higher efficiency.
Table~\ref{tab:poly_relu_comparison} illustrates
the approximation polynomial with different degrees.
For degree $6$, the polynomial is $0.1249 + 0.5000x + 0.3729x^2 -0.0410x^4 + 0.0016x^6$.
Computing it takes $13$ multiplications and $4$ additions, which translates to $82.20$ms.
Our ReLU circuits thus outperform polynomial approximation by $265$ times.

\subsubsection{Evaluation on MNIST and CIFAR-10 dataset}
Table~\ref{tab:eachphase} reports the latency for each stage of the network 
on the MNIST dataset 
(consists of $9$ stages as presented in Fig.~\ref{fig:neural_network_architecture_on_MNIST})
and CIFAR-10 dataset 
(consists of $17$ stages as described in Fig.~\ref{fig:neural_network_architecture_on_CIFAR10}).
In the offline phase, the two servers interact with each other to prepare the triplets and the ReLU circuits, which costs $0.471$s and $7.942$s on the MNIST dataset and CIFAR-10 dataset.
In the online phase, the activation function dominates since it uses garbled circuits.
For the convolution layer, pooling layer, and fully connected layer, all the computations are executed over shares, which just takes a little time.
Notably, for the client, encoding the query and decoding the prediction result just cost $5.568$\textmu s and $21.818$\textmu s on the MNIST dataset and CIFAR-10 dataset, respectively.

\begin{table*}[!ht]
\setlength{\belowcaptionskip}{3pt}
\centering
\newcolumntype{M}[1]{>{\centering\arraybackslash}m{#1}}
\newcolumntype{R}[1]{>{\raggedleft\arraybackslash}m{#1}}
\renewcommand{\arraystretch}{1.6}
\addtolength{\tabcolsep}{-0pt}
\caption{Performance of the synchronous and asynchronous computations on the MNIST and CIFAR-10 datasets}
\begin{tabular}{ M{8em}||M{3.8em}|M{3.8em}||M{3.8em}|M{3.8em}||M{3.8em}|M{3.8em} ||M{3.8em}|M{3.8em}||M{3.8em}|M{3.8em}}
 \Xhline{1.1pt}
 \multirow{3}{*}{}	&	\multicolumn{8}{c||}{MNIST dataset}	&	\multicolumn{2}{c}{CIFAR-10 dataset}\\
\cline{2-11}
	&	\multicolumn{2}{c||}{Network 1}	&	\multicolumn{2}{c||}{Network 2}	&	\multicolumn{2}{c||}{Network 3}	&	\multicolumn{2}{c||}{Network 4}	&	\multicolumn{2}{c}{Network 5} \\
\cline{2-11}
	&	Syn. Comp	&	Asyn. Comp	&	Syn. Comp	&	Asyn. Comp	&	Syn. Comp	&	Asyn. Comp	&	Syn. Comp	&	Asyn. Comp	&	Syn. Comp	&	Asyn. Comp\\
\Xhline{1.1pt}
 Offline Phase (s)	&	$0.007$	&	$0.005$	&	$0.020$	&	$0.016$	&	$0.052$	&	$0.050$	&	$0.471$	&	$0.469$	&	$7.946$	&	$7.845$\\
\hline
Online Phase (s)	&	$0.033$	&	$0.023$	&	$0.050$	&	$0.039$	&	$0.323$	&	$0.319$	&	$3.363$	&	$3.234$	&	$68.278$	&	$65.553$\\
\hline
Total (s)	&	$0.040$	&	$0.028$	&	$0.070$	&	$0.057$	&	$0.375$	&	$0.368$	&	$3.835$	&	$3.703$	&	$76.224$	&	$73.398$\\
\Xhline{1.1pt}
\end{tabular}
\label{tab:PerformanceAsynSyn}
\end{table*}

\begin{figure*}[!ht]
\centering
\begin{minipage}[!htbp]{0.48\linewidth}
\centering
\includegraphics[width=0.99\textwidth]{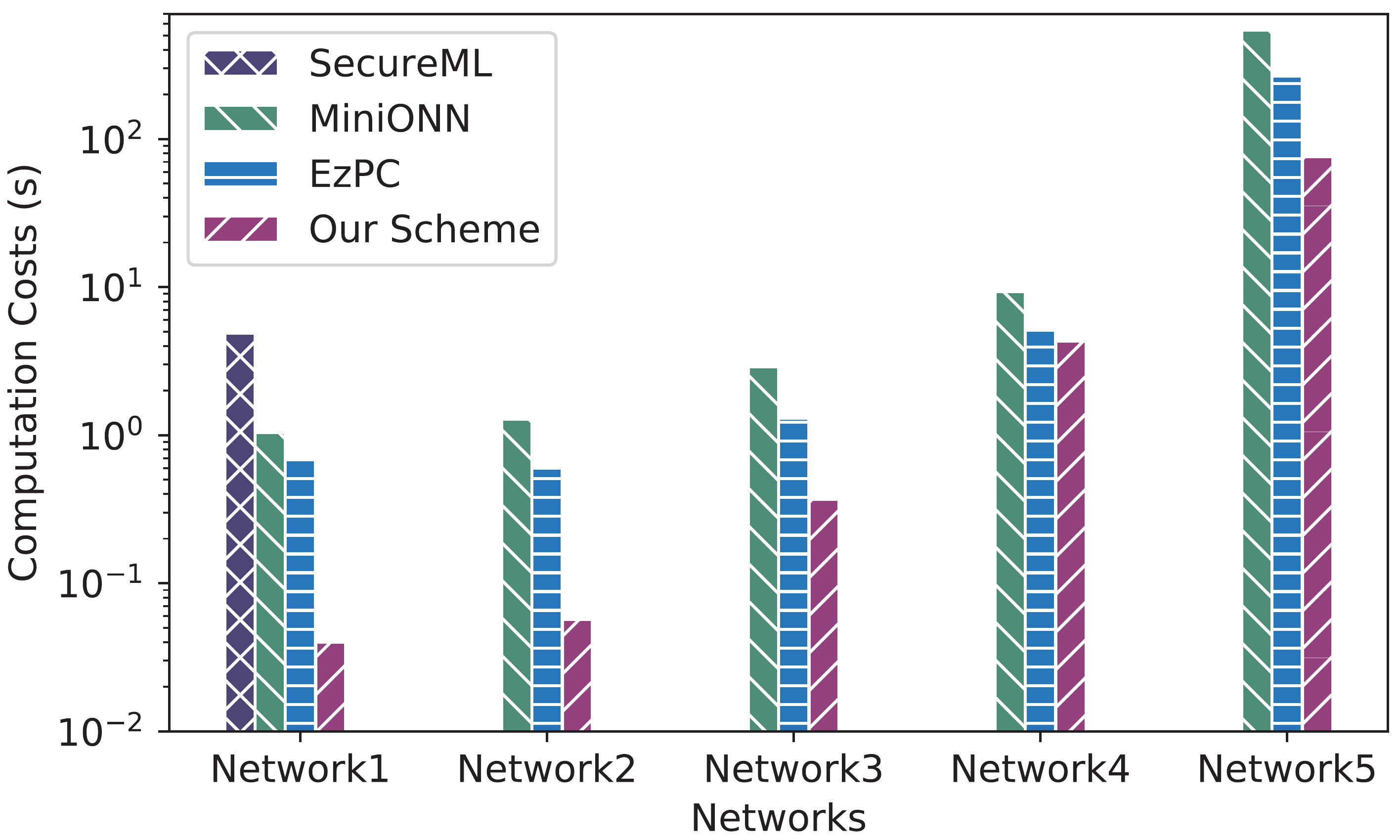}
\caption{Comparison of the computation overheads}
\label{fig:comp_costs}
\end{minipage}%
\hspace{10pt}
\begin{minipage}[!htbp]{0.48\linewidth}
\centering
\includegraphics[width=0.99\textwidth]{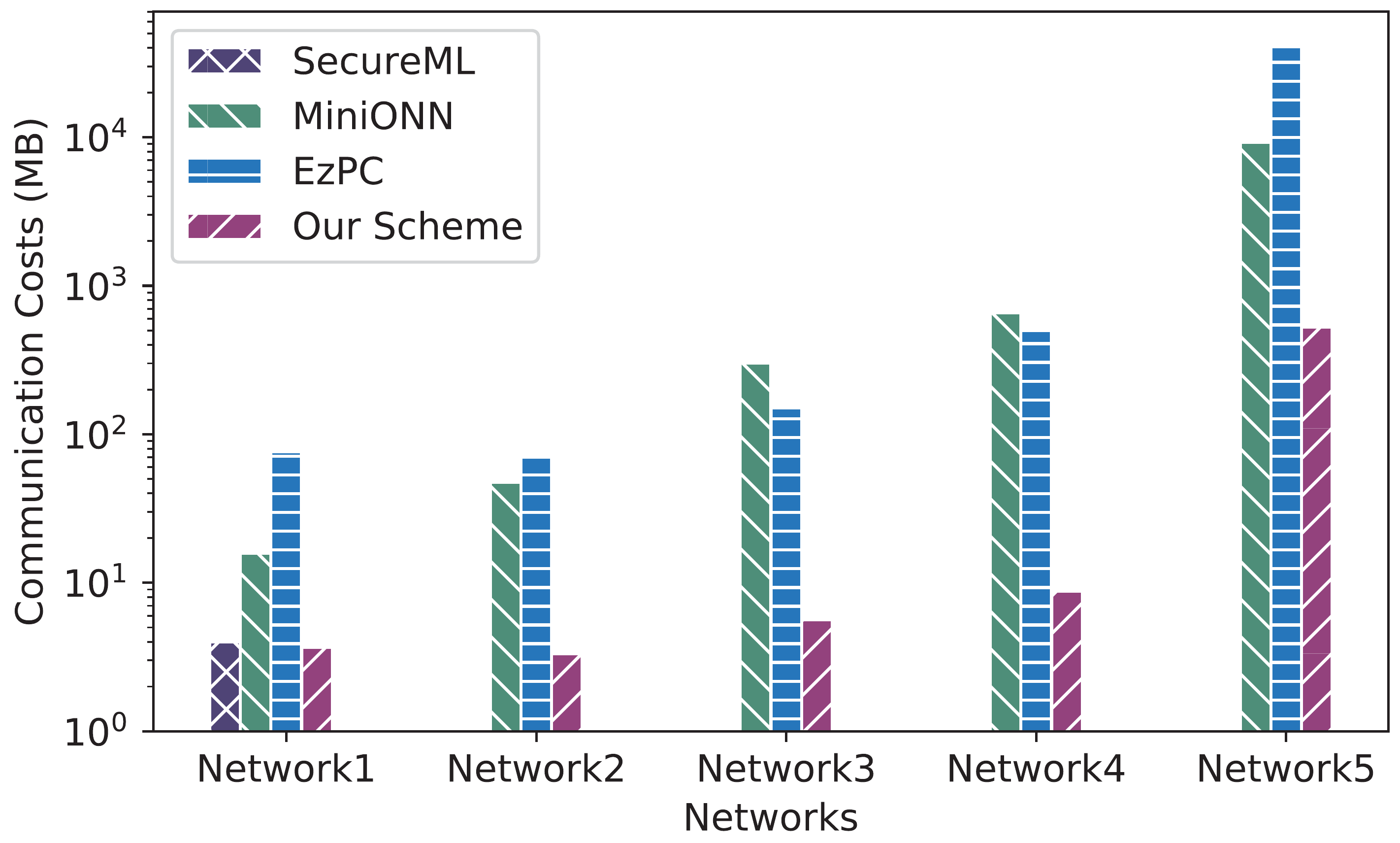}
\caption{Comparison of the communication overheads}
\label{fig:comm_costs}
\end{minipage}
\end{figure*}

\subsubsection{Evaluation on Different Network Architectures}
We conduct experiment over network architectures in five published works~\cite{MohasselZ17,dac/RouhaniRK18,Gilad-BachrachD16,LiuJLA17},
which use a combination of FC and Conv layers as follows.
For FC$(784\rightarrow 128)$, $784$ and $128$ respectively represent the size of the input and the output.
On the other hand,
for Conv$(1\times28\times28\rightarrow 5\times13\times13)$, the input image is of size $28\times28$ with $1$ channel, the output image is of size $13\times13$ with $5$ channels.
For the square activation function, it is essentially multiplication operation between two secret shares.
We can achieve this by using the triplet similar to the convolutional operation in our scheme.
\begin{itemize}
 \item Network1~\cite{MohasselZ17}: FC$(784\rightarrow 128)$ $\Rightarrow$ Square $\Rightarrow$ FC$(128\rightarrow128)$ $\Rightarrow$ Square $\Rightarrow$ FC$(128\rightarrow10)$.
 \item Network2~\cite{dac/RouhaniRK18}: Conv$(1\times28\times28\rightarrow5\times13\times13)$ $\Rightarrow$ ReLU $\Rightarrow$ FC$(845\rightarrow100)$ $\Rightarrow$ ReLU $\Rightarrow$ FC$(100\rightarrow10)$.
 \item Network3~\cite{Gilad-BachrachD16}: Conv$(1\times28\times28\rightarrow5\times13\times13)$ $\Rightarrow$ Square $\Rightarrow$ Pooling$(5\times13\times13\rightarrow5\times13\times13)$ $\Rightarrow$ Conv $(5\times13\times13\rightarrow50\times5\times5)$ $\Rightarrow$ Pooling$(50\times5\times5\rightarrow50\times5\times5)$ $\Rightarrow$ FC$(1250\rightarrow100)$ $\Rightarrow$ Square $\Rightarrow$ FC$(100\rightarrow10)$.
 \item Network4~\cite{LiuJLA17}: the same as in Fig.~\ref{fig:neural_network_architecture_on_MNIST}.
 \item Network5~\cite{LiuJLA17}: the same as in Fig.~\ref{fig:neural_network_architecture_on_CIFAR10}.
\end{itemize}

Table~\ref{tab:PerformanceAsynSyn} shows the performance of synchronous and asynchronous computations in both the offline and online phases on the five networks.
For network1, the total time cost of the existing schemes with synchronous computation is $0.040$s,
while the time cost of our scheme with asynchronous computation is just $0.028$s.
In other words, we obtain a $30.0$ \% performance saving.
Analogously, the savings for network2, network3, network4, and network5 are $18.57\%$, $1.87\%$, $3.44\%$, and $3.71\%$, respectively.
In short, our asynchronous computation reduces latency.

To demonstrate the superiority of our scheme, we
compare with
SecureML~\cite{MohasselZ17}, MiniONN~\cite{LiuJLA17}, and EzPC~\cite{eurosp/ChandranGRST19}.
Fig.~\ref{fig:comp_costs} and Fig.~\ref{fig:comm_costs} show the computation and communication costs on different neural network architectures, respectively.

On the MNIST dataset (with Network1, 2, 3, 4), compared with MiniONN~\cite{LiuJLA17} (without model privacy), our scheme performs an average of $14.63\times$ and $36.69\times$ improvements on computation and communication costs, respectively.
Compared with EzPC~\cite{eurosp/ChandranGRST19} (without model privacy), ours is $8.19\times$ and $31.32\times$ better on average (of Network1, 2, 3, 4) for the computation and communication costs, respectively.
Compared with SecureML~\cite{MohasselZ17} (with model privacy) using Network1, we provide $122\times$ faster computation time and $1.09\times$ lower communication cost.

On the CIFAR dataset using Network5, our scheme achieves $7.14\times$ and $3.48\times$ improvements in reducing latency compared to MiniONN and EzPC.
For the communication costs, our scheme outperforms MiniONN by $13.88\times$, and EzPC by $77.46\times$ times.

\section{Security Analysis}

Both the query and the model are randomly shared and distributed to two independent servers.
As long as they do not collude, the privacy of the query and the model are preserved.
More specifically, both servers always process them either over secret shares or over garbled circuits for non-linear activation,
with 
the help of pre-computed triplets, which are independent of the sensitive data.
So, no meaningful information is revealed to either of the servers.

Below we provide a simulation-based proof for security.

\textit{Security Definition}: Let $f=(f_{\ds}, f_{\sone}, f_{\stwo})$ be a probabilistic polynomial function and $\prod$ a protocol computing $f$.
The user and two servers want to compute $f(\mathcal{G}, \mathsf{C}^1, \mathsf{C}^2, \mathsf{W}^1, \mathsf{W}^2)$ using $\prod$ where $\mathcal{G}$ is the query image of the user $\ds$,
and $(\mathsf{C}^1, \mathsf{W}^1)$ and $(\mathsf{C}^2, \mathsf{W}^2)$
are the shared model deposited on the two servers.
The view of $\ds$ during the execution of $\prod$ is
$V_{\ds}(\mathcal{G}, \mathsf{C}^1, \mathsf{C}^2, \mathsf{W}^1, \mathsf{W}^2) = (\mathcal{G}, r_{\ds}, m_{\ds})$ where $r_{\ds}$ is the random tape of $\ds$, and $m_U$ is the message received by $\ds$.
Simultaneously, the view of $\sone$ and $\stwo$ are defined as $V_{\sone}(\mathcal{G}, \mathsf{C}^1, \mathsf{C}^2, \mathsf{W}^1, \mathsf{W}^2) = ( \mathsf{C}^1, \mathsf{W}^1, r_{\sone}, m_{\sone})$, $V_{\stwo}(a, b, c) =$ $( \mathsf{C}^2, \mathsf{W}^2, r_{\stwo}, m_{\stwo})$.
The protocol $\prod$ achieving the function $f$ is regarded as secure if for every possible input $\mathcal{G}, \mathsf{C}^1, \mathsf{C}^2, \mathsf{W}^1, \mathsf{W}^2$ of $f$, there exist the probabilistic polynomial time simulators $\Phi_{\ds}$, $\Phi_{\sone}$, and $\Phi_{\stwo}$ such that,
\begin{align*}
\Phi_{\ds}(a, f_{\ds}(\mathcal{G}, \mathsf{C}^1, \mathsf{C}^2, \mathsf{W}^1, \mathsf{W}^2))		&	\equiv_{p} V_{\ds} (\mathcal{G}, \mathsf{C}^1, \mathsf{C}^2, \mathsf{W}^1, \mathsf{W}^2),\\
\Phi_{\sone}(a, f_{\sone}(\mathcal{G}, \mathsf{C}^1, \mathsf{C}^2, \mathsf{W}^1, \mathsf{W}^2))	&	\equiv_{p} V_{\sone} (\mathcal{G}, \mathsf{C}^1, \mathsf{C}^2, \mathsf{W}^1, \mathsf{W}^2),\\
\Phi_{\stwo}(a, f_{\stwo}(\mathcal{G}, \mathsf{C}^1, \mathsf{C}^2, \mathsf{W}^1, \mathsf{W}^2))	&	\equiv_{p} V_{\stwo} (\mathcal{G}, \mathsf{C}^1, \mathsf{C}^2, \mathsf{W}^1, \mathsf{W}^2),
\end{align*}
where $\equiv_{p}$ denotes computational indistinguishability.

\vspace{3pt}
\textit{Theorem 1: The triplet generation protocol in Section~\ref{subsec:triplet_generation} is secure against
semi-honest adversaries.}

\textit{Proof}:
$\sone$ holds the secret key.
All the messages passed from $\sone$ to $\stwo$ are encrypted.
All
those
passed from $\stwo$ to $\sone$ are distributed uniformly by adding
random data
independent of the data of $\stwo$.
The view of $\sone$ is $V_{\sone} =(\mathsf{C}^1, \mathsf{W}^1, a_1, b_1, z_1)$,
where $a_1, b_1, z_1$ come from the triplet of $\sone$.
We
construct
the simulator $\Phi_{\sone}(\mathsf{C}^1, \mathsf{W}^1)$ as:

1) Pick the random integers $\hat{a}_1, \hat{b}_1,\hat{ z}_1$ from $\mathbb{Z}_{2^t}$.

2) Output $\Phi_{\sone}(\mathsf{C}^1, \mathsf{W}^1)=(\mathsf{C}^1, \mathsf{W}^1, \hat{a}_1, \hat{b}_1,\hat{ z}_1)$.

As the randomness $\hat{a}_1$ (or $\hat{b}_1, \hat{z}_1$) is generated in the same manner as $a_1$ (or $a_1, b_1$), and independent from the other data, the distribution of $(\mathsf{C}^1, \mathsf{W}^1, \hat{a}_1, \hat{b}_1,\hat{ z}_1)$ and $(\mathsf{C}^1, \mathsf{W}^1, {a}_1, {b}_1,{ z}_1)$ are indistinguishable.
Thus we
have proven that $V_{\sone} \equiv_{p} \Phi_{\sone}(\mathsf{C}^1, \mathsf{W}^1)$.
Analogously, $\Phi_{\stwo}(\mathsf{C}^2, \mathsf{W}^2) =$ $(\mathsf{C}^2, \mathsf{W}^2, \hat{a}_2, \hat{b}_2,\hat{ z}_2)$, $V_{\stwo} \equiv_{p} \Phi_{\stwo}(\mathsf{C}^2, \mathsf{W}^2)$.

\vspace{3pt}
\textit{Theorem 2: The query distribution protocol in Section~\ref{subsec:query_distribution} is secure against
semi-honest adversaries.}

\textit{Proof}:
User $\ds$ additively shares the query image into two parts.
The views of the three parties are $V_{\ds} =(\mathcal{G}, R_{\mathcal{G}})$, $V_{\sone}=(\mathsf{C}^1, \mathsf{W}^1, \mathcal{G}^1)$, $V_{\stwo}= (\mathsf{C}^2, \mathsf{W}^2, \mathcal{G}^2)$, where $\mathcal{G}^1= R_{\mathcal{G}}$, $\mathcal{G}^2=\mathcal{G} - R_{\mathcal{G}}$.
We construct a simulator $\Phi_{\sone}(\mathsf{C}^1, \mathsf{W}^1)$ as:

1) Pick random integers $\hat{R}_{\mathcal{G}}$ from $\mathbb{Z}_{2^t}$ and set $\hat{\mathcal{G}}^1= \hat{R}_{\mathcal{G}}$.

2) Output $\Phi_{\sone}(\mathsf{C}^1, \mathsf{W}^1)=(\mathsf{C}^1, \mathsf{W}^1, \hat{\mathcal{G}}^1)$.

Both $R_{\mathcal{G}}$ and $\hat{R}_{\mathcal{G}}$ are generated randomly.
Hence the distribution of $R_{\mathcal{G}}$ and $\hat{R}_{\mathcal{G}}$ are indistinguishable.
As a consequence, we have $V_{\sone} \equiv_{p} \Phi_{\sone}(\mathsf{C}^1, \mathsf{W}^1)$.
Analogously, $\Phi_{\stwo}(\mathsf{C}^2, \mathsf{W}^2) = (\mathsf{C}^2, \mathsf{W}^2, \hat{\mathcal{G}}^2)$, $V_{\stwo} \equiv_{p} \Phi_{\stwo}(\mathsf{C}^2, \mathsf{W}^2)$.

\vspace{3pt}
\textit{Theorem 3: The convolutional computation protocol in Section~\ref{subsec:conv_comp} is secure against
semi-honest adversaries.}

\textit{Proof}:
The two servers
perform the convolution
with the help of triplets.
Their views $V_{\sone}$ and $V_{\stwo}$ are
$(\mathsf{C}^1, \mathsf{W}^1, \mathcal{G}^1, \mathcal{H}^1, A_1, B_1, Z_1)$,
$(\mathsf{C}^2, \mathsf{W}^2, \mathcal{G}^2, \mathcal{H}^2, A_1, B_1, Z_1)$
respectively.
We construct a simulator $\Phi_{\sone}(\mathsf{C}^1, \mathsf{W}^1)$ as:

1) Call the triplet generation protocol to have $\hat{A}_1, \hat{B}_1, \hat{Z}_1$.

2) Pick the random integers $\hat{\mathcal{G}}^1, \hat{U}^2, \hat{V}^2$ from $\mathbb{Z}_{2^t}$.

3) Compute $\hat{U}=\hat{\mathcal{G}}^1 - \hat{A}_1 + \hat{U}^2$, $\hat{V}=\mathsf{C}^1 - \hat{B}_1 + \hat{V}^2$.

4) Compute $\hat{\mathcal{H}}^1 = - \hat{U}\hat{V} + \hat{\mathcal{G}}^1\hat{V} + \mathsf{C}^1\hat{U} + \hat{Z}_1$.

5) Output $\Phi_{\sone}(\mathsf{C}^1, \mathsf{W}^1)=(\mathsf{C}^1, \mathsf{W}^1, \hat{\mathcal{G}}^1, \hat{\mathcal{H}}^1)$.

The distribution of $\mathcal{H}^1$ and $\hat{\mathcal{H}}^1$ are indistinguishable.
Hence $V_{\sone} \equiv_{p} \Phi_{\sone}(\mathsf{C}^1, \mathsf{W}^1)$ holds.
Analogously, we have $\Phi_{\stwo}(\mathsf{C}^2, \mathsf{W}^2)=(\mathsf{C}^2, \mathsf{W}^2, \hat{\mathcal{G}}^2, \hat{\mathcal{H}}^2)$, and $V_{\stwo} \equiv_{p} \Phi_{\stwo}(\mathsf{C}^2, \mathsf{W}^2)$.

\vspace{3pt}
\textit{Theorem 4: The activation computation protocol in Section~\ref{subsec:activation_comp} is secure against
semi-honest adversaries.}

\textit{Proof}:
The two servers compute the activation function by using the designed garbled circuits.
The view of $\sone$ is $V_{\sone}=(\mathsf{C}^1, \mathsf{W}^1, \mathcal{H}^1, \mathcal{J}^1)$. The view of $\stwo$ is $V_{\stwo}=(\mathsf{C}^2, \mathsf{W}^2, \mathcal{H}^2, \mathcal{J}^2)$.
Since the garbled circuits are secure against semi-honest adversaries,
we only consider the inputs and outputs of the circuits rather than the internal details.
We construct a simulator $\Phi_{\sone}(\mathsf{C}^1, \mathsf{W}^1)$ as:

1) Pick the random integers $\hat{\mathcal{H}}^1, \hat{\mathcal{J}}^1$ from $\mathbb{Z}_{2^t}$.

2) Output $\Phi_{\sone}(\mathsf{C}^1, \mathsf{W}^1)=(\mathsf{C}^1, \mathsf{W}^1, \hat{\mathcal{H}}^1, \hat{\mathcal{J}}^1)$.

The distributions of $\mathcal{H}^1$ ($\mathcal{J}^1$) and $\mathcal{H}^1$ ($\hat{\mathcal{J}}^1$) are indistinguishable.
Hence $V_{\sone} \equiv_{p} \Phi_{\sone}(\mathsf{C}^1, \mathsf{W}^1)$ holds.
Analogously, we have $\Phi_{\stwo}(\mathsf{C}^2, \mathsf{W}^2)=(\mathsf{C}^2, \mathsf{W}^2, \hat{\mathcal{H}}^2, \hat{\mathcal{J}}^2)$, and $V_{\stwo} \equiv_{p} \Phi_{\stwo}(\mathsf{C}^2, \mathsf{W}^2)$.

\vspace{3pt}
\textit{Theorem 5: The pooling computation protocol in Section~\ref{subsec:pooling_comp} is secure against
semi-honest adversaries.}

\textit{Proof}:
The view of $\sone$ is $V_{\sone}=(\mathsf{C}^1, \mathsf{W}^1, \mathcal{J}^1, \mathcal{K}^1)$. The view of $\stwo$ is $V_{\stwo}=(\mathsf{C}^2, \mathsf{W}^2, \mathcal{J}^2, \mathcal{K}^2)$.
We construct a simulator $\Phi_{\sone}(\mathsf{C}^1, \mathsf{W}^1)$ as:

1) Pick the random integers $\hat{\mathcal{J}}^1, \hat{\mathcal{K}}^1$ from $\mathbb{Z}_{2^t}$.

2) Output $\Phi_{\sone}(\mathsf{C}^1, \mathsf{W}^1)=(\mathsf{C}^1, \mathsf{W}^1, \hat{\mathcal{J}}^1, \hat{\mathcal{K}}^1)$.

Since the distribution of $\mathcal{J}^1$ ($\mathcal{K}^1$) and $\mathcal{J}^1$ ($\hat{\mathcal{K}}^1$) are indistinguishable, $V_{\sone} \equiv_{p} \Phi_{\sone}(\mathsf{C}^1, \mathsf{W}^1)$ holds.
Analogously,
$\Phi_{\stwo}(\mathsf{C}^2, \mathsf{W}^2)=(\mathsf{C}^2, \mathsf{W}^2, \hat{\mathcal{J}}^2, \hat{\mathcal{K}}^2)$, and $V_{\stwo} \equiv_{p} $ $\Phi_{\stwo}(\mathsf{C}^2, \mathsf{W}^2)$.

\vspace{3pt}
\textit{Theorem 6: The fully connected protocol in Section~\ref{subsec:fully_connected} is secure against
semi-honest adversaries.}

\textit{Proof}:
The views $V_{\sone}$ and $V_{\stwo}$
are $(\mathsf{C}^1, \mathsf{W}^1, \mathcal{K}^1, \mathcal{F}^1, A_1, B_1, Z_1)$,
$(\mathsf{C}^2, \mathsf{W}^2, \mathcal{K}^2, \mathcal{F}^2, A_2, B_2, Z_2)$ respectively.
We construct a simulator $\Phi_{\sone}(\mathsf{C}^1, \mathsf{W}^1)$ as:

1) Call the triplet generation protocol to have $\hat{A}_1, \hat{B}_1, \hat{Z}_1$.

2) Pick the random integers $\hat{\mathcal{K}}^1, \hat{U}^2, \hat{V}^2$ from $\mathbb{Z}_{2^t}$.

3) Compute $\hat{U}=\hat{\mathcal{K}}^1 - \hat{A}_1 + \hat{U}^2$, $\hat{V}=\mathsf{W}^1 - \hat{B}_1 + \hat{V}^2$.

4) Compute $\hat{\mathcal{F}}^1 = - \hat{U}\hat{V} + \hat{\mathcal{K}}^1\hat{V} + \mathsf{W}^1\hat{U} + \hat{Z}_1$.

5) Output $\Phi_{\sone}(\mathsf{C}^1, \mathsf{W}^1)=(\mathsf{C}^1, \mathsf{W}^1, \hat{\mathcal{K}}^1, \hat{\mathcal{F}}^1)$.

The distribution of $\mathcal{F}^1$ and $\hat{\mathcal{F}}^1$ are indistinguishable.
Hence $V_{\sone} \equiv_{p} \Phi_{\sone}(\mathsf{C}^1, \mathsf{W}^1)$ holds.
Analogously, we have $\Phi_{\stwo}(\mathsf{C}^2, \mathsf{W}^2)=(\mathsf{C}^2, \mathsf{W}^2, \hat{\mathcal{K}}^2, \hat{\mathcal{F}}^2)$, and $V_{\stwo} \equiv_{p} \Phi_{\stwo}(\mathsf{C}^2, \mathsf{W}^2)$.

\section{Conclusion and Future Work}
Neural-network prediction has unprecedented accuracy in many tasks, notably in image-based algorithms.
However, privacy concerns have worried people when they hand in their data as a query.
On the other hand, we see a growing market for machine learning model, 
the worry of model privacy is more pressing.
We improve neural-network prediction in the secure outsourcing setting
The privacy guarantees is that the servers providing the prediction service can get nothing about the query, the model, any intermediate results, and the final result.
We design garbled circuits for non-linear activation function, which preserves the accuracy of the underlying neural network.
We also reduce the overheads of computation and communication by adopting packing and asynchronous computation.
Our experiments over both MNIST and CIFAR-10 datasets showcase our improvement.

As a very active research area, many works consider additional features.
We discuss two limitations of our approach.
Our scheme fails to provide verifiability, \ie, the servers can deviate from the protocol specification and return wrong results. 
In critical applications, it is necessary to consider such possibility of malicious acts.
Fortunately, the non-colluding assumption could be ``fully leveraged'' for achieving verifiability, specifically, by designing the corresponding maliciously-secure protocols, assuming more than two non-colluding servers~\cite{ccs/MohasselR18}.

Another limitation of our result
is that we did not exploit GPU at all, despite its popularity in plaintext machine-learning. 
Some recent works integrate both SGX and GPU for prediction~\cite{iclr/TramerB19} and training~\cite{ng2020goten}.
Using the GPU to aid secure computation with SGX for training is already highly non-trivial~\cite{iclr/TramerB19,ng2020goten},
using it to further improve cryptographic approaches would be a very interesting and challenging future direction.

\bibliographystyle{IEEEtran}
\bibliography{reference}

\begin{thebibliography}{10}
\providecommand{\url}[1]{#1}
\csname url@samestyle\endcsname
\providecommand{\newblock}{\relax}
\providecommand{\bibinfo}[2]{#2}
\providecommand{\BIBentrySTDinterwordspacing}{\spaceskip=0pt\relax}
\providecommand{\BIBentryALTinterwordstretchfactor}{4}
\providecommand{\BIBentryALTinterwordspacing}{\spaceskip=\fontdimen2\font plus
\BIBentryALTinterwordstretchfactor\fontdimen3\font minus
  \fontdimen4\font\relax}
\providecommand{\BIBforeignlanguage}[2]{{%
\expandafter\ifx\csname l@#1\endcsname\relax
\typeout{** WARNING: IEEEtran.bst: No hyphenation pattern has been}%
\typeout{** loaded for the language `#1'. Using the pattern for}%
\typeout{** the default language instead.}%
\else
\language=\csname l@#1\endcsname
\fi
#2}}
\providecommand{\BIBdecl}{\relax}
\BIBdecl

\bibitem{Goodfellow-et-al-2016}
I.~Goodfellow, Y.~Bengio, and A.~Courville, \emph{Deep Learning}.\hskip 1em
  plus 0.5em minus 0.4em\relax MIT Press, 2016,
  \url{http://www.deeplearningbook.org}.

\bibitem{tifs/ZhaoWZZC19}
L.~Zhao, Q.~Wang, Q.~Zou, Y.~Zhang, and Y.~Chen, ``Privacy-preserving
  collaborative deep learning with unreliable participants,'' \emph{{IEEE}
  Trans. Information Forensics and Security}, vol.~15, pp. 1486--1500, 2020.

\bibitem{tifs/ZouWWZL20}
Q.~Zou, Y.~Wang, Q.~Wang, Y.~Zhao, and Q.~Li, ``Deep learning-based gait
  recognition using smartphones in the wild,'' \emph{{IEEE} Trans. Information
  Forensics and Security}, vol.~15, pp. 3197--3212, 2020.

\bibitem{tvt/ZouJDYCW20}
Q.~Zou, H.~Jiang, Q.~Dai, Y.~Yue, L.~Chen, and Q.~Wang, ``Robust lane detection
  from continuous driving scenes using deep neural networks,'' \emph{{IEEE}
  Trans. Veh. Technol.}, vol.~69, no.~1, pp. 41--54, 2020.

\bibitem{uss/TramerZJRR16}
F.~Tram{\`{e}}r, F.~Zhang, A.~Juels, M.~K. Reiter, and T.~Ristenpart,
  ``Stealing machine learning models via prediction {APIs},'' in \emph{USENIX
  Security Symposium}, 2016, pp. 601--618.

\bibitem{uss/AdiBCPK18}
Y.~Adi, C.~Baum, M.~Ciss{\'{e}}, B.~Pinkas, and J.~Keshet, ``Turning your
  weakness into a strength: Watermarking deep neural networks by backdooring,''
  in \emph{USENIX Security Symposium}, 2018, pp. 1615--1631.

\bibitem{sp/ShokriSSS17}
R.~Shokri, M.~Stronati, C.~Song, and V.~Shmatikov, ``Membership inference
  attacks against machine learning models,'' in \emph{IEEE Symposium on
  Security and Privacy ({S\&P})}, 2017, pp. 3--18.

\bibitem{popets/HayesMDC19}
J.~Hayes, L.~Melis, G.~Danezis, and E.~D. Cristofaro, ``{LOGAN:} membership
  inference attacks against generative models,'' \emph{PoPETs}, vol. 2019,
  no.~1, pp. 133--152, 2019.

\bibitem{ndss/Salem0HBF019}
A.~Salem, Y.~Zhang, M.~Humbert, P.~Berrang, M.~Fritz, and M.~Backes,
  ``{ML-Leaks}: Model and data independent membership inference attacks and
  defenses on machine learning models,'' in \emph{ISOC Network and Distributed
  System Security Symposium (NDSS)}, 2019.

\bibitem{ccs/PapernotMGJCS17}
N.~Papernot, P.~D. McDaniel, I.~J. Goodfellow, S.~Jha, Z.~B. Celik, and
  A.~Swami, ``Practical black-box attacks against machine learning,'' in
  \emph{{ACM} Asia Conference on Computer and Communications Security
  (AsiaCCS)}, 2017, pp. 506--519.

\bibitem{uss/JiaG18}
J.~Jia and N.~Z. Gong, ``{AttriGuard}: {A} practical defense against attribute
  inference attacks via adversarial machine learning,'' in \emph{USENIX
  Security Symposium}, 2018, pp. 513--529.

\bibitem{nips/ElsayedSCPKGS18}
G.~F. Elsayed, S.~Shankar, B.~Cheung, N.~Papernot, A.~Kurakin, I.~J.
  Goodfellow, and J.~Sohl{-}Dickstein, ``Adversarial examples that fool both
  computer vision and time-limited humans,'' in \emph{Neural Information
  Processing Systems (NeurIPS)}, 2018, pp. 3914--3924.

\bibitem{tdsc/WangSZZSW19}
Z.~Wang, M.~Song, S.~Zheng, Z.~Zhang, Y.~Song, and Q.~Wang, ``Invisible
  adversarial attack against deep neural networks: An adaptive penalization
  approach,'' \emph{{IEEE} Trans. Dependable Sec. Comput.}, 2019, {DOI}:
  10.1109/TDSC.2019.2929047.

\bibitem{AbadiCGMMT016}
M.~Abadi, A.~Chu, I.~J. Goodfellow, H.~B. McMahan, I.~Mironov, K.~Talwar, and
  L.~Zhang, ``Deep learning with differential privacy,'' in \emph{ACM
  Conference on Computer and Communications Security (CCS)}, 2016, pp.
  308--318.

\bibitem{SmithTU17}
A.~D. Smith, A.~Thakurta, and J.~Upadhyay, ``Is interaction necessary for
  distributed private learning?'' in \emph{IEEE Symposium on Security and
  Privacy ({S\&P})}, 2017, pp. 58--77.

\bibitem{OhrimenkoSFMNVC16}
O.~Ohrimenko, F.~Schuster, C.~Fournet, A.~Mehta, S.~Nowozin, K.~Vaswani, and
  M.~Costa, ``Oblivious multi-party machine learning on trusted processors,''
  in \emph{USENIX Security Symposium}, 2016, pp. 619--636.

\bibitem{tdsc/HuZWQW19}
S.~Hu, L.~Y. Zhang, Q.~Wang, Z.~Qin, and C.~Wang, ``Towards private and
  scalable cross-media retrieval,'' \emph{{IEEE} Trans. Dependable Sec.
  Comput.}, 2019, {DOI}: 10.1109/TDSC.2019.2926968.

\bibitem{iclr/TramerB19}
F.~Tram{\`{e}}r and D.~Boneh, ``Slalom: Fast, verifiable and private execution
  of neural networks in trusted hardware,'' in \emph{International Conference
  on Learning Representations (ICLR)}, 2019.

\bibitem{scc/Chow19}
S.~S.~M. Chow, ``Can we securely outsource big data analytics with lightweight
  cryptography?'' in \emph{International Workshop on Security in Cloud
  Computing, SCC@ASIACCS}, 2019, p.~1.

\bibitem{ng2020goten}
L.~K.~L. Ng, S.~S.~M. Chow, A.~P.~Y. Woo, D.~P.~H. Wong, and Y.~Zhao, ``Goten:
  {GPU}-outsourcing trusted execution of neural network training and
  prediction,'' 2019.

\bibitem{Gilad-BachrachD16}
R.~Gilad{-}Bachrach, N.~Dowlin, K.~Laine, K.~E. Lauter, M.~Naehrig, and
  J.~Wernsing, ``{CryptoNets}: Applying neural networks to encrypted data with
  high throughput and accuracy,'' in \emph{International Conference on Machine
  Learning (ICML)}.\hskip 1em plus 0.5em minus 0.4em\relax ACM, 2016, pp.
  201--210.

\bibitem{ChabanneWMMP17}
H.~Chabanne, A.~de~Wargny, J.~Milgram, C.~Morel, and E.~Prouff,
  ``Privacy-preserving classification on deep neural network,'' {IACR}
  Cryptology ePrint Archive 2017/035, 2017.

\bibitem{LiuJLA17}
J.~Liu, M.~Juuti, Y.~Lu, and N.~Asokan, ``Oblivious neural network predictions
  via {MiniONN} transformations,'' in \emph{ACM Conference on Computer and
  Communications Security (CCS)}, 2017, pp. 619--631.

\bibitem{uss/JuvekarVC18}
C.~Juvekar, V.~Vaikuntanathan, and A.~Chandrakasan, ``{GAZELLE:} {A} low
  latency framework for secure neural network inference,'' in \emph{USENIX
  Security Symposium}, 2018, pp. 1651--1669.

\bibitem{dac/RouhaniRK18}
B.~D. Rouhani, M.~S. Riazi, and F.~Koushanfar, ``{DeepSecure}: scalable
  provably-secure deep learning,'' in \emph{Design Automation Conference
  (DAC)}, 2018, pp. 2:1--2:6.

\bibitem{uss/RiaziSCLLK19}
M.~S. Riazi, M.~Samragh, H.~Chen, K.~Laine, K.~E. Lauter, and F.~Koushanfar,
  ``{XONN:} {XNOR}-based oblivious deep neural network inference,'' in
  \emph{USENIX Security Symposium}, 2019, pp. 1501--1518.

\bibitem{eurosp/ChandranGRST19}
N.~Chandran, D.~Gupta, A.~Rastogi, R.~Sharma, and S.~Tripathi, ``{EzPC}:
  Programmable, efficient, and scalable secure two-party computation,'' in
  \emph{EuroS{\&}P}.\hskip 1em plus 0.5em minus 0.4em\relax {IEEE}, 2019, pp.
  496--511.

\bibitem{gentry09}
C.~Gentry, ``Fully homomorphic encryption using ideal lattices,'' in
  \emph{Symposium on Theory of Computing (STOC)}.\hskip 1em plus 0.5em minus
  0.4em\relax ACM, 2009, pp. 169--178.

\bibitem{ndss/BostPTG15}
R.~Bost, R.~A. Popa, S.~Tu, and S.~Goldwasser, ``Machine learning
  classification over encrypted data,'' in \emph{ISOC Network and Distributed
  System Security Symposium (NDSS)}, 2015.

\bibitem{esorics17/TaiMZC}
R.~K.~H. Tai, J.~P.~K. Ma, Y.~Zhao, and S.~S.~M. Chow, ``Privacy-preserving
  decision trees evaluation via linear functions,'' in \emph{European Symposium
  on Research in Computer Security (ESORICS)}, 2017, pp. 494--512.

\bibitem{tdsc/AloufiHWC19}
A.~Aloufi, P.~Hu, H.~W.~H. Wong, and S.~S.~M. Chow, ``Blindfolded evaluation of
  random forests with multi-key homomorphic encryption,'' \emph{{IEEE} Trans.
  Dependable Sec. Comput.}, 2019, {DOI}: 10.1109/TDSC.2019.2940020.

\newpage


\bibitem{fcs/Chow18}
S.~S.~M. Chow, ``Privacy-preserving machine learning ({Invited paper}),'' in
  \emph{Frontiers in Cyber Security. FCS 2018}, ser. Communications in Computer
  and Information Science, vol. 879.\hskip 1em plus 0.5em minus 0.4em\relax
  Springer, 2018.

\bibitem{HesamifardTGW18}
E.~Hesamifard, H.~Takabi, M.~Ghasemi, and R.~N. Wright, ``Privacy-preserving
  machine learning as a service,'' \emph{PoPETs}, vol. 2018, no.~3, pp.
  123--142, 2018.

\bibitem{ccs/JiangKLS18}
X.~Jiang, M.~Kim, K.~E. Lauter, and Y.~Song, ``Secure outsourced matrix
  computation and application to neural networks,'' in \emph{ACM Conference on
  Computer and Communications Security (CCS)}, 2018, pp. 1209--1222.

\bibitem{ndss/ChowLS09}
S.~S.~M. Chow, J.~Lee, and L.~Subramanian, ``Two-party computation model for
  privacy-preserving queries over distributed databases,'' in \emph{ISOC
  Network and Distributed System Security Symposium (NDSS)}, 2009.

\bibitem{iacr/KamaraMR11}
S.~Kamara, P.~Mohassel, and M.~Raykova, ``Outsourcing multi-party
  computation,'' {IACR} Cryptology ePrint Archive 2011/272.

\bibitem{cns/WangLC014}
B.~Wang, M.~Li, S.~S.~M. Chow, and H.~Li, ``A tale of two clouds: Computing on
  data encrypted under multiple keys,'' in \emph{{IEEE} Conference on
  Communications and Network Security, {CNS}}, 2014, pp. 337--345.

\bibitem{BaryalaiJL16}
M.~Baryalai, J.~Jang{-}Jaccard, and D.~Liu, ``Towards privacy-preserving
  classification in neural networks,'' in \emph{Privacy, Security and Trust
  (PST)}, 2016, pp. 392--399.

\bibitem{tkde/WangDCCZCH18}
Q.~Wang, M.~Du, X.~Chen, Y.~Chen, P.~Zhou, X.~Chen, and X.~Huang,
  ``Privacy-preserving collaborative model learning: The case of word vector
  training,'' \emph{{IEEE} Trans. Knowl. Data Eng.}, vol.~30, no.~12, pp.
  2381--2393, 2018.

\bibitem{MohasselZ17}
P.~Mohassel and Y.~Zhang, ``{SecureML}: {A} system for scalable
  privacy-preserving machine learning,'' in \emph{IEEE Symposium on Security
  and Privacy ({S\&P})}.\hskip 1em plus 0.5em minus 0.4em\relax IEEE, 2017, pp.
  19--38.

\bibitem{cacm/Shamir79}
A.~Shamir, ``How to share a secret,'' \emph{Commun. {ACM}}, vol.~22, no.~11,
  pp. 612--613, 1979.

\bibitem{Beaver91a}
D.~Beaver, ``Efficient multiparty protocols using circuit randomization,'' in
  \emph{Advances in Cryptology -- CRYPTO}, 1991, pp. 420--432.

\bibitem{Yao86}
A.~C. Yao, ``How to generate and exchange secrets,'' in \emph{IEEE Symposium on
  Foundations of Computer Science}.\hskip 1em plus 0.5em minus 0.4em\relax
  IEEE, 1986, pp. 162--167.

\bibitem{dcc/SmartV14}
N.~P. Smart and F.~Vercauteren, ``Fully homomorphic {SIMD} operations,''
  \emph{Des. Codes Cryptography}, vol.~71, no.~1, pp. 57--81, 2014.

\bibitem{ndss/Demmler0Z15}
D.~Demmler, T.~Schneider, and M.~Zohner, ``{ABY} - {A} framework for efficient
  mixed-protocol secure two-party computation,'' in \emph{ISOC Network and
  Distributed System Security Symposium (NDSS)}, 2015.

\bibitem{ima/BosLLN13}
J.~W. Bos, K.~E. Lauter, J.~Loftus, and M.~Naehrig, ``Improved security for a
  ring-based fully homomorphic encryption scheme,'' in \emph{IMA Int'l
  Conference on Cryptography and Coding (IMACC)}, 2013, pp. 45--64.

\bibitem{pieee/DowlinGLLNW17}
N.~Dowlin, R.~Gilad{-}Bachrach, K.~Laine, K.~E. Lauter, M.~Naehrig, and
  J.~Wernsing, ``Manual for using homomorphic encryption for bioinformatics,''
  \emph{Proc. of the {IEEE}}, vol. 105, no.~3, pp. 552--567, 2017.

\bibitem{lecun1998mnist}
Y.~LeCun, ``The {MNIST} database of handwritten digits,''
  \url{http://yann.lecun.com/exdb/mnist}, 1998.

\bibitem{KrizhevskyH09}
A.~Krizhevsky and G.~Hinton, ``Learning multiple layers of features from tiny
  images,'' Tech. Rep., 2009.

\bibitem{ccs/MohasselR18}
P.~Mohassel and P.~Rindal, ``{ABY}\({}^{\mbox{3}}\): {A} mixed protocol
  framework for machine learning,'' in \emph{ACM Conference on Computer and
  Communications Security (CCS)}.\hskip 1em plus 0.5em minus 0.4em\relax {ACM},
  2018, pp. 35--52.

\end{thebibliography}

\pretolerance=-1
\tolerance=10000
\emergencystretch=100em
\hbadness=1000
\hfuzz=1pt

\newpage
\appendix
\section*{Further Illustration}\label{app:figure}
\begin{figure}[h]
\centering
\includegraphics[width=1\columnwidth]{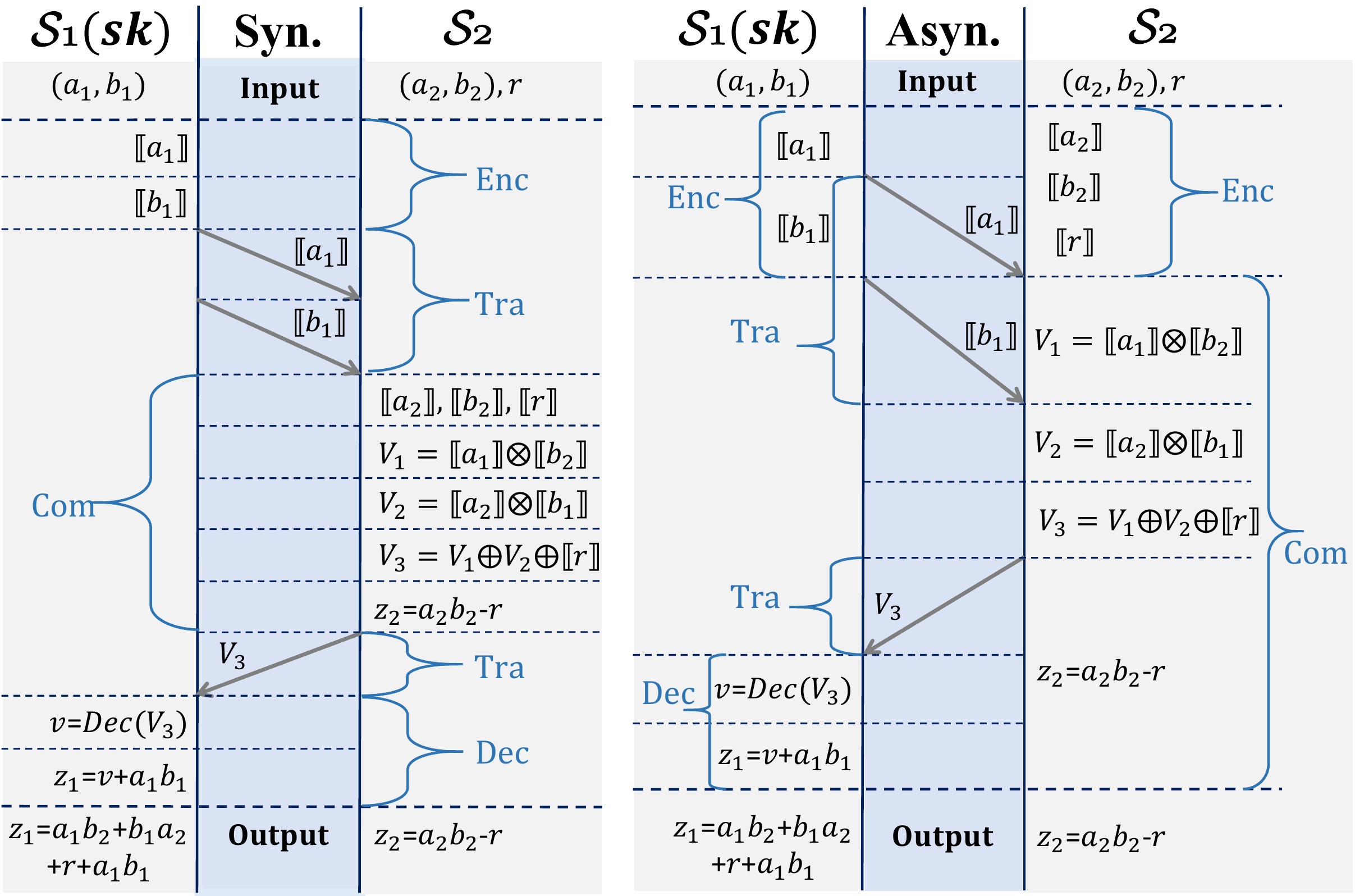}
\caption{Synchronous/Asynchronous computations}
\label{fig:sync_async_comp}
\end{figure}

\begin{figure}[h]
\centering
\includegraphics[width=1\columnwidth]{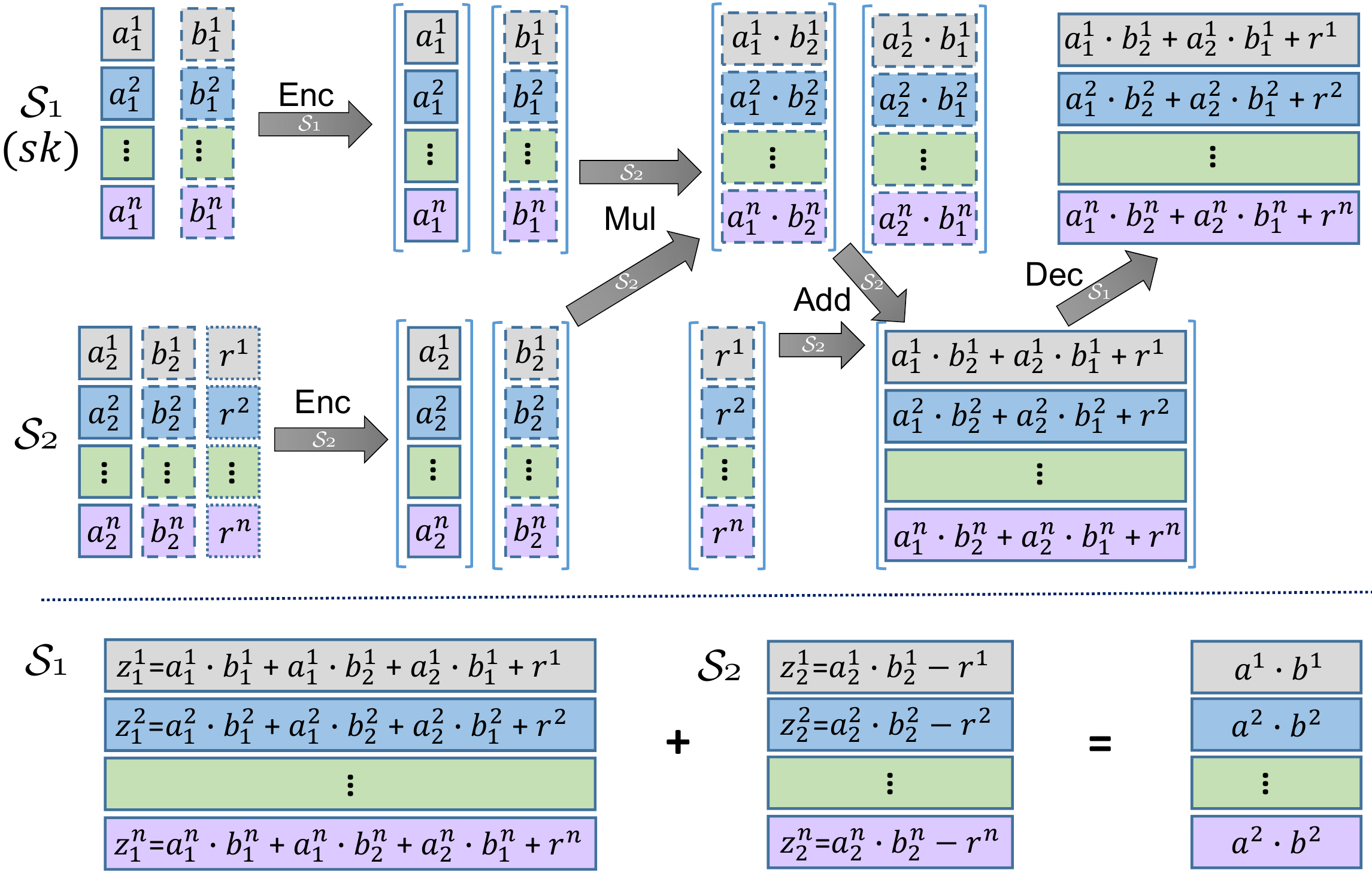}
\caption{A toy example of simultaneous triplet generation for $4$ triplets}
\label{fig:simd_example}
\end{figure}

Here we provide two supplementary illustrations for our triplet generation protocol in the offline phase.
Fig.~\ref{fig:sync_async_comp} (left) illustrates the existing synchronous design.
Our asynchronous design, which keeps the servers busy with the remaining operations that do not involve feedback from the other, 
is illustrated in Fig.~\ref{fig:sync_async_comp} (right)
Finally, Fig.~\ref{fig:simd_example} illustrates a toy example for the packed triplet generation.

\end{document}